\documentclass[
					aps, 
					pre,								
					reprint,     
					showpacs, 					
					preprintnumbers,		
					amsmath,
					amssymb,
 					english
					]{revtex4-1}

\usepackage{bm}	
\usepackage[english]{babel}

\begin{document}

\preprint{Os.R52/12-en77}

\title{Open statistical ensemble and surface phenomena} 

\author{V. M. Zaskulnikov}
\homepage{http://www.zaskulnikov.ru}
\email[]{zaskulnikov@gmail.com }
\affiliation{Novosibirsk, Russia}

\date{\today}

\begin{abstract}

In the present work we investigate a new statistical ensemble, which seems logical to be entitled the open one, for the case of a one-component system of ordinary particles. Its peculiarity is in complementing the consideration of a system with the inclusion of a certain surrounding area. The calculations indicate the necessity of taking into account the surface that delimits a given system even in the case when the latter is a part of a uniform medium and is not singled out one way or another.

The surface tension coefficient behaves unlike two-phase systems in equilibrium and depends on two variables - pressure as well as temperature - and belongs to the boundary separating a hard solid from a fluid. As for the mathematical mechanism ensuring the fulfillment of thermodynamic relations, the emphasis is shifted from operating with series, like in the grand canonical ensemble, towards employing the recurrence relations of a new class of functions that incorporate Boltzmann and Ursell factors as their extreme cases and towards utilizing generating functions. 

The second topic of discussion that the present article deals with is the consideration of the surface tension and adsorption observed at the boundary of a solid body and a liquid or gas carried out on the basis of the analysis of the classical system found in a field of force of general type. The surface terms are calculated with the aid of field functions and the correlation functions of an unperturbed volume phase and behave somewhat vaguely; particularly, as a function of activity, they may start with a linear or quadratic term.

\end{abstract}


\maketitle

\section{\label{sec:01}Introduction}

The Gibbs's grand canonical distribution is one of the cornerstones of statistical physics and its basic part leaves no place for doubt. Nevertheless, it seems worthwhile to check how well the relations of the grand canonical ensemble (GCE) meet certain criteria. One of those criteria is a specific kind of conformity that consists in a subsystem distribution (calculated on the basis of the GCE ) matching the initial one. 

Right from the beginning, the situation asks for such a test. Indeed, a well-known study of nucleation that can be traced back to J.W. Gibbs \cite[p.242]{gibbs1961} does examine the surface of a nucleus. However the original grand canonical distribution describes fluctuations of the particle number but at first sight does not seem to include any terms that represent surface. 

The incorporation of the terms describing the interaction over a surrounding surface into the statistical distribution requires considering the environment that lies outside the limits of the system under analysis. This brings forward the idea of the open statistical ensemble (OSE). It can be further developed in the course of carrying out the conformity check. 

Take note that in this case surface terms emerge when we describe a uniform medium that has no real surface but only a hypothetical one setting bounds to the system under consideration. This unexpected result can be better understood if we pay attention to the fact that the point of interest for us in statistical distributions is the fluctuations occurring strictly inside the limits of a singled out volume. The states corresponding to these fluctuations already contain the delimiting surface which causes the abovementioned effect. Surface terms show the largest values for large fluctuations and are balanced out for mean values. It is also quite curious that a surface term shall be part of a standard GCE distribution, though with some deformations.  

With the treatment of the OSE distribution as a fluctuation probability, coherence between the thermodynamic and statistical approaches is established and the need for including surface terms into the distribution is justified.

In an ordinary case of a two-phase one-component system the surface tension coefficient depends on one variable only - either pressure or temperature - due to the existence of the phase equilibrium curve. The situation under consideration extends beyond the reach of this condition and the surface tension coefficient starts depending on two variables. As it will be later shown it corresponds to the surface tension observed in the situation when a fluid interacts with a hard solid.

Lately, the notion of the surface tension existing at the boundary between a solid body and a fluid has been on the agenda. For instance, it has been developing under the framework of density functional theory. The present article, however, suggests an alternative approach based on investigating a classical system situated in a field of force of general type which yields the combination of the field functions (solid body) and the correlation functions of an unperturbed fluid array. 

This approach produces the results that coincide with the direction initiated in \cite{Bellemans1962}; however, it is significantly simpler than the latter one allowing us to write the general form for the expansion in powers of activity.

As a result, we've obtained the expression for $\Omega$-potential that gives us, in the limits of smooth fields, logically justified thermodynamic expressions for a nonuniform system. For the fields that are microscopically inhomogeneous we obtain the surface terms corresponding, for the potential of a hard solid, to the coefficient that is part of the open ensemble partition function. For realistic potentials the volume term, surface tension and adsorption are not likely to be subject to strict separation within the limits of a transition layer. 

Such treatment, as a matter of fact, deals with an insoluble (nonvaporizing) solid body, adjacent to a liquid or gas.  It seems only logical that in the setting of nonequilibrium the equilibrium curve does not have impact on the result.

\section{\label{sec:02}Cavity Formation Probability}

This quantity is a vital parameter that defines the partition function of an ensemble. Thus, we will start from dealing with the first and most important term of this distribution and then we will switch to other ones. Let us begin with the canonical ensemble, move forward to the GCE and eventually come up with the definition for the OSE. 

Consider the probability of a fluctuation hole formation inside a uniform statistical system, that is an area with the volume $v$ that is devoid of any particles. In order to facilitate the calculations we will, in the majority of cases, have the integration volume set by the window functions $\psi^v_i$ that are determined by the following relation: 
\begin{equation}
\psi^v(\bm{r}_i) = \psi^v_i = 
	 \left\{ 
			\begin{array}{ll} 
         1 & (\bm{r}_i \in v)\\   
         0 & (\bm{r}_i \notin v),
     	\end{array}  
		\right.
		\label{eq:001}
\end{equation}
where $v$ indicates the area volume while the integration will be considered to be performed over infinite space unless stated otherwise.

We should also note that the algebra of window functions plays a significant role in the approach being described. Apart from that, as it will be further seen, the importance of these functions is high from the point of view of determining correlation functions.

\subsection{\label{subsec:02a}Canonical Ensemble}

The required probability in an isolated system shall evidently be defined as
\begin{equation}
p^v_{C,0} = \frac{1}{Z_N}\int \left [ \prod_{k = 1}^N(\psi^V_k-\psi^v_k) \right ] \exp(-\beta U^N_{1...N}) d\bm{r}_1...d\bm{r}_N.
\label{eq:002}  
\end{equation}

Here $N$ is the number of particles in the system, $\beta = 1/k_BT$, $k_B$ - Boltzmann constant, $T$ - temperature, $U^N_{1...N}$ is the energy of particle interaction, and $V$  is the system volume. $C$ at $p^v_{C,0}$ indicates that the probability is defined for the canonical ensemble - on the condition of finding $N$ particles inside volume $V$. $Z_N$ is the configuration integral:
\begin{equation}
Z_N = \int \left [ \prod_{k = 1}^N \psi^V_k \right ]  \exp(-\beta U^N_{1...N}) d\bm{r}_1...d\bm{r}_N. 
\label{eq:003}
\end{equation}

By expanding the product in (\ref{eq:002}) and separating the integration over $v$, we will have:
\begin{equation}
p^v_{C,0} = 1 + \sum_{k=1}^N \binom{N}{k} (-1)^k \int \left [ \prod_{l = 1}^k \psi^v_l \right ]  \ P^{(k)}_{1...k} d\bm{r}_1...d\bm{r}_k,
\label{eq:004}
\end{equation}
where  $P^{(k)}_{1...k}$ has the form:
\begin{equation}
P^{(k)}_{1...k} = \frac{1}{Z_N}\int \left [ \prod_{l = k+1}^N \psi^V_l \right ] \exp(-\beta U^N_{1...N})d\bm{r}_{k+1}...d\bm{r}_N. 
\label{eq:005}
\end{equation}

$P^{(k)}_{1...k}$ defines the density of probability to find a given space arrangement of a set of specific particles  \cite[p.181]{hillstatmeh1987}. Considering distribution functions for a  set of arbitrary particles 
\begin{equation}
\varrho^{(k)}_{C,1...k} = \frac{N!}{(N-k)!}P^{(k)}_{1...k}, 
\label{eq:006}
\end{equation}
from (\ref{eq:004}) we shall have:
\begin{equation}
p^v_{C,0} = 1 + \sum_{k=1}^N \frac{(-1)^k}{k!}  \int \left [ \prod_{l = 1}^k \psi^v_l \right ] \varrho^{(k)}_{C,1...k} d\bm{r}_1...d\bm{r}_k.
\label{eq:007}
\end{equation}

The analysis of a canonical ensemble causes some inconvenience due to distorted asymptotic forms of the functions $\varrho^{(k)}_{C,1...k}$ \cite[p.408]{hillstatmeh1987}, which results in integrals becoming non-local. Thus, a shift towards a GCE is required.

\subsection{\label{subsec:02b}Grand Canonical Ensemble}

Let us average the Eq. (\ref{eq:007}) over the fluctuations of the number of particles by applying to both of its sides the operation $\sum_{N=0}^\infty P_N$, where 
\begin{equation}
P_N = \frac{z^N Z_N}{N!\Xi_V} 
\label{eq:008}
\end{equation}
- is the probability that the GCE shall have a definite number of particles $N$ inside region $V$. Here $z$ - is activity:
\begin{equation}
z = \frac{e^{\mu/k_BT}}{\Lambda^3}, 
\label{eq:009}
\end{equation}
where $\mu$ - is chemical potential, $\Lambda = h/\sqrt[]{2 \pi mk_BT}$, $h$ - Planck constant, $m$ - particle mass and $\Xi_V$ - is the large partition function of a system with the volume $V$:
\begin{equation}
\Xi_V = \sum_{N=0}^\infty \frac{z^N Z_N}{N!}. 
\label{eq:010}
\end{equation}

During the averaging we may change the order of summing in the right-hand side (\ref{eq:007}) which yields

\begin{equation}
p^v_{G,0} = 1 + \sum_{k=1}^\infty \frac{(-1)^k}{k!}  \int \left [ \prod_{l = 1}^k \psi^v_l \right ] \varrho^{(k)}_{G,1...k}(\psi^V) d\bm{r}_1 ...d\bm{r}_k.
\label{eq:011}
\end{equation}
Here
\begin{equation}
\varrho^{(k)}_{G,1...k}(\psi^V) = \sum_{N=k}^\infty \varrho^{(k)}_{C,1...k} P_N,
\label{eq:012}
\end{equation}
or in another way 
\begin{eqnarray}
\varrho^{(k)}_{G,1...k}(\psi^V) &=& \frac{z^k}{\Xi_V}\sum_{N=0}^\infty \frac{z^N}{N!} \int \left [ \prod_{l = k+1}^{k+N} \psi^V_l \right ]  \label{eq:013} \\
&\times&   \exp(-\beta U^{N+k}_{1...N+k})d\bm{r}_{k+1}...d\bm{r}_{k+N}.  \nonumber
\end{eqnarray}

$\varrho^{(k)}_{G,1...k}(\psi^V)$ - is the $k$- partical distribution function for a GCE. As stated above, these functions set the density of probability to find a certain configuration of arbitrary particles. For an ideal gas $\varrho^{(k)}_{G,1...k} = \varrho^{k}$, where $\varrho = \overline{N}/V$ - is the density of particles. 

$\psi^V$ cannot be regarded as a range of definition of functions in a general sense of the term, since, as we will further see, coordinates of free particles may extend beyond the boundaries that it sets. For the sake of conciseness, we will be calling it a range of assignment. For instance, in Eq. (\ref{eq:013}) $\psi^V$ is the range of assignment of $\varrho^{(k)}_{G,1...k}$. 

In this case these functions make their entry through the series in the numerator (\ref{eq:013}) as well as in the denominator - through the series (\ref{eq:010}) for $\Xi_V$. Further on, it will be analyzed more in detail.

Further investigation will reveal how significant the range of assignment $\psi^V$  is, and for this reason we will be explicitly denoting it like we do with the belonging to the GCE-type by indicating  it with $G$. 

Usually, the range of assignment $\psi^V$ is not paid enough attention, thus being barely registered and weakly controlled, but according to the treatment that this article offers it is vital.

Let us take the logarithm of the Eq. (\ref{eq:011}) in order to localize the integrals.

\begin{equation}
\ln{p^v_{G,0}} = \sum_{k=1}^\infty \frac{(-1)^k}{k!}  \int \left [ \prod_{l = 1}^k \psi^v_l \right ] {\cal F}^{(k)}_{G,1...k}(\psi^V) d\bm{r}_1...d\bm{r}_k.
\label{eq:014}
\end{equation}

Here ${\cal F}^{(k)}_{G,1...k}(\psi^V)$ - are Ursell functions or localized correlation functions \cite{ursell1927}. We will be also pointing out their range of assignment and type affiliation whenever it is necessary.

Some relations will be true both for functions of the GCE-type and the OSE-type ones. In these cases, we will be leaving out the attributes of the GCE to avoid the exceeding sophistication of formulas.

We shall now consider the properties of correlation functions and factors before moving any further.

\section{\label{sec:03}Correlation Functions}

We shall further need the correlation functions of three types: complete, localized and partially localized ones as well as the three corresponding types of factors.

Aside from that, the correlation functions will vary according to the type of the range of assignment.

\subsection{\label{subsec:03a}Complete Correlation Functions}

To draw more certainty to the definition this term will be applied to functions $\varrho^{(k)}_{G,1...k}(\psi^V)$, determined by Eq. (\ref{eq:013}) and to their contiguous functions from the OSE - $\varrho^{(k)}_{1...k}$, which will be defined further.

We will use their following property: 
\begin{eqnarray}
\int  \psi^V_k{\varrho}^{(k)}_{G,1...k} \mkern -7mu &(& \mkern -7mu \psi^V)d\bm{r}_k \label{eq:015} \\
&=& \left[\frac{\partial}{\partial \beta \mu } + \overline{N} - k + 1 \right]{\varrho}^{(k-1)}_{G,1...k-1}(\psi^V).
\nonumber
\end{eqnarray}

This recurrence relation can be easily obtained from, for instance, (\ref{eq:012}) by differentiation performed with respect to chemical potential \footnote[1]{All differentiations are hereinafter performed with constant volume and temperature values, unless it is stated otherwise.}.

We shall also note the following property of their complete integrals:
\begin{equation}
\int \left [ \prod_{i = 1}^k \psi^V_i \right ] {\varrho}^{(k)}_{G,1...k} (\psi^V) d\bm{r}_1... d\bm{r}_k = \left \langle  \frac{N!}{(N-k)!} \right \rangle,
\label{eq:016}
\end{equation}
which follows from the definition.

Equations (\ref{eq:015}) and (\ref{eq:016}) are precise; however, the very values $\langle{N}\rangle$ for the GCE, as we shall witness further, incorporate inaccuracies connected to surface terms.

An important property of these functions is the split into a product as two groups of particles are drifting apart:
\begin{equation}
\varrho^{(k + l)}_{1...k + l} \to \varrho^{(k)}_{1...k} \varrho^{(l)}_{k + 1...k + l}.
\label{eq:017}
\end{equation}

\subsection{\label{subsec:03b}Localized Correlation Functions}

Localized correlation functions or Ursell functions (they are also termed connected, inner and truncated correlation functions) have been analyzed by many authors and, particularly, in publications \cite{MayerGeppert1977, percus1964, LebowitzPercus1961, duneau1973}.

These functions exhibit the local property for they quickly die away as any groups of particles, including a single particle only, start distancing from each other. This is observed due to property (\ref{eq:017}). Let us write some of the first functions for a uniform system.
\begin{eqnarray}
{\cal F}^{(1)}_{1}\mkern 15mu & = & \varrho\label{eq:018}  \\
{\cal F}^{(2)}_{1,2} \mkern 16mu& = & \varrho^{(2)}_{1,2} -  \varrho^2 \nonumber \\
{\cal F}^{(3)}_{1,2,3}\mkern 10mu & = & \varrho^{(3)}_{1,2,3} -  ( \varrho^{(2)}_{1,2} + \varrho^{(2)}_{2,3} + \varrho^{(2)}_{1,3})\varrho  + 2\varrho^{3} \nonumber \\
{\cal F}^{(4)}_{1,2,3,4} & = & \varrho^{(4)}_{1,2,3,4} -  (\varrho^{(3)}_{1,2,3} + \varrho^{(3)}_{2,3,4} + \varrho^{(3)}_{1,3,4} + \varrho^{(3)}_{1,2,4})\varrho\nonumber \\
& - & (\varrho^{(2)}_{1,2}\varrho^{(2)}_{3,4}  +  \varrho^{(2)}_{2,3}\varrho^{(2)}_{1,4} + \varrho^{(2)}_{1,3}\varrho^{(2)}_{2,4})+ 2(\varrho^{(2)}_{1,2} + \varrho^{(2)}_{2,3}\nonumber \\
& + & \varrho^{(2)}_{1,3}  +  \varrho^{(2)}_{1,4}  +  \varrho^{(2)}_{2,4} + \varrho^{(2)}_{3,4})\varrho^2 - 6\varrho^{4}
\nonumber \\
\dotso \nonumber 
\end{eqnarray}

For an arbitrary $k \geq 1$, we shall have \cite[Chapter 9]{MayerGeppert1977}:
\begin{eqnarray}
{\cal F}^{(k)}_{1...k} &=& \sum_{\{\bm{n}\}}(-1)^{l-1}(l-1)!\prod_{\alpha = 1}^l \varrho^{(k_\alpha)}(\{\bm{n}_\alpha\}) \nonumber \\
1 & \leq & k_\alpha \leq k ~~~~~~~ \sum_{\alpha = 1}^l k_\alpha = k, \label{eq:019}
\end{eqnarray}
where $\{\bm{n}\}$ stands for a certain decomposition of a given set of $k$ particles located at $\bm{r}_1,...\bm{r}_k$ into non-overlapping groups $\{\bm{n}_\alpha\}$, with $l$ being the number of groups of particular partition, $k_\alpha$ - the size of group number $\alpha$, with the summing performed over all possible divisions. 

An inverse relation of the following type is correct:
\begin{equation}
\varrho^{(k)}_{1...k} = \sum_{\{\bm{n}\}}\prod_{\alpha = 1}^l {\cal F}^{(k_\alpha)}(\{\bm{n}_\alpha\}).
\label{eq:020}
\end{equation}

The (\ref{eq:011}) $\leftrightarrow$ (\ref{eq:014}) type of relation is universal. If 
\begin{equation}
\mu(\phi) = 1 + \sum_{k=1}^\infty \frac{1}{k!}  \int \left [ \prod_{l = 1}^k \phi_l \right ] \mu^{(k)}_{1...k} d\bm{r}_1 ...d\bm{r}_k
\label{eq:021}
\end{equation}
and
\begin{equation}
\nu(\phi) = \ln [\mu(\phi)],
\label{eq:022}
\end{equation}
then
\begin{equation}
\nu(\phi) = \sum_{k=1}^\infty \frac{1}{k!}  \int \left [ \prod_{l = 1}^k \phi_l \right ] \nu^{(k)}_{1...k} d\bm{r}_1...d\bm{r}_k, 
\label{eq:023}
\end{equation}
where  $\mu^{(k)}_{1...k}$ and $\nu^{(k)}_{1...k}$ are sets of functions bound by the relations of types (\ref{eq:019}), (\ref{eq:020}) \cite{percus1964}. 

For ${\cal F}^{(k)}_{G,1...k}(\psi^V)$ the following recurrence relation is true: 
\begin{equation}
\int\psi^V_k {\cal F}^{(k)}_{G,1...k}(\psi^V) d\bm{r}_k = \left[\frac{\partial}{\partial \beta \mu } - k + 1\right]{\cal F}^{(k-1)}_{G,1...k-1}(\psi^V).
\label{eq:024}
\end{equation}

Obviously, (\ref{eq:024}) can be obtained from (\ref{eq:015}) on the basis of definition (\ref{eq:019}). The fact that $\langle N \rangle$ vanishes from (\ref{eq:024}) as opposed to Eq. (\ref{eq:015}) reflects the local property of these functions indeed. Eq. (\ref{eq:024}) is equivalent to 
\begin{equation}
\int\psi^V_k {\cal F}^{(k)}_{G,1...k}(\psi^V) d\bm{r}_k = z^k  \frac{\partial}{\partial z } \left[ \frac{{\cal F}^{(k-1)}_{G,1...k-1}(\psi^V)}{z^{k-1}} \right].
\label{eq:025}
\end{equation}

Multiple integration leads to: 
\begin{eqnarray}
\int \left [ \prod_{i = l}^k \psi^V_i \right ] {\cal F}^{(k)}_{G,1...k}(\psi^V\mkern -7mu&)&\mkern -7mu d\bm{r}_l...d\bm{r}_k\label{eq:026} \\
 &=& z^k  \frac{\partial^{k-l+1}}{\partial z^{k-l+1} } \left[ \frac{{\cal F}^{(l-1)}_{G,1...l-1}(\psi^V)}{z^{l-1}} \right], \nonumber
\end{eqnarray}
where $l=2,3...k$. For $l=2$ we shall write
\begin{eqnarray}
\int\left [ \prod_{i = 2}^k \psi^V_i \right ] {\cal F}^{(k)}_{G,1...k} (\psi^V \mkern -7mu&)&\mkern -7mu d\bm{r}_2...d\bm{r}_k \label{eq:027} \\
&=& z^k  \frac{\partial^{k-1}}{\partial z^{k-1} } \left[ \frac{\varrho_G(\psi^V)}{z} \right].
\nonumber
\end{eqnarray}

Taking into consideration that
\begin{equation}
\varrho=\left(\frac{\partial P}{\partial \mu} \right)_{T} = z\beta \left(\frac{\partial P}{\partial z} \right)_{T},
\label{eq:028}
\end{equation} 
we eventually have 
\begin{equation}
\int\left [ \prod_{i = 2}^k \psi^V_i \right ] {\cal F}^{(k)}_{G,1...k}(\psi^V) d\bm{r}_2...d\bm{r}_k \approx z^k \beta \frac{\partial^{k}P}{\partial z^{k} }.
\label{eq:029}
\end{equation}

Eq. (\ref{eq:029}), that initially appeared in \cite{duneau1973} during the investigation of the analytical properties of $P(z)$, is, generally speaking, approximate because $\varrho_G$ does not correspond to $\varrho$ in the vicinity of the boundaries of volume. This question will be studied further on.

We shall also note that integrals (\ref{eq:027}) and (\ref{eq:029}), as a consequence of the local property of inner correlations, are defined by the microscopic region of integration in the vicinity of the first particle, unless the rank of the functions enters the macroscopic domain $k \sim 10^{23}$.

These are equations (\ref{eq:024}) - (\ref{eq:027}), (\ref{eq:029}) that are analogous to Ornstein-Zernike relations for the GCE rather than the integrals found in \cite{LebowitzPercus1961}.

\subsection{\label{subsec:03c}Partially Localized Correlation Functions}

These functions constitute a hybrid of full and localized correlation functions. They play an important role in the mathematical apparatus of the OSE. As far as we are concerned, they were first introduced in work \cite{duneau1973} for the expansion of complete correlation functions. We will further see, though, that it is more convenient to employ the correlation functions normalized by $z^{m+k}$, where $m+k$ is their total rank.

Some of the particles that are part of these functions behave like those belonging to a complete correlation function, i.e. they do not cause dying out as they move away, whereas the other part shows local properties.

Let us set partially localized correlation functions as:
\begin{equation}
{\cal A}^{(m,k)}_{1...m+k},
\label{eq:030}
\end{equation}
where superscripts $m$ and $k$ set the number of delocalized and localized particles correspondingly ($m = 1,2,3,\dots, k = 0,1,2,\dots$). In their turn, the subscripts denote the coordinates of the particles with the restriction that the first $m$ particles are treated as delocalized, and the rest of them as those possessing local properties. 

The structure of these functions is analogous to that of localized correlation functions of the $(k+1)$-th rank, on the condition that the first $m$ particles (the delocalized ones) are considered as one compound particle when writing (\ref{eq:019}).

In other words:
\begin{eqnarray}
{\cal A}^{(m,k)}_{1...m+k} &=& \frac{1}{z^{m+k}}\sum_{\{\bm{n}\}}(-1)^{l-1}\label{eq:031}\\
&\times&(l-1)!\prod_{\alpha = 1}^l \varrho^{(k_\alpha+(m-1)\delta_{\alpha\gamma})}(\{\bm{n}_\alpha\}) \nonumber \\
&1 & \leq ~ k_\alpha \leq k+1 ~~~~~ \sum_{\alpha = 1}^l k_\alpha = k+1, \nonumber
\end{eqnarray}
where the denotations are analogous to those from (\ref{eq:019}) on the condition that the summation is performed over all possible decompositions of $k+1$ particles, one of which is compound. $\delta_{\alpha\gamma}$ is the Kronecker delta, and $\gamma$ is the number of the group containing the compound particle.

We may further write some of the first functions (over a local grouping) for a uniform system:
\begin{eqnarray}
 {\cal A}^{(m,0)}_{1...m} \mkern 12mu & = & \varrho^{(m)}_{1...m}/z^m   \label{eq:032}  \\
 {\cal A}^{(m,1)}_{1...m+1} & = & [\varrho^{(m+1)}_{1...m+1} -  \varrho^{(m)}_{1...m} \varrho]/z^{m+1} \nonumber \\
 {\cal A}^{(m,2)}_{1...m+2} & = & [\varrho^{(m+2)}_{1...m+2} -  \varrho^{(m+1)}_{1...m+1}\varrho - \varrho^{(m+1)}_{1...m,m+2}\varrho \nonumber \\
 &-& \varrho^{(m)}_{1...m}\varrho^{(2)}_{m+1,m+2}  + 2\varrho^{(m)}_{1...m}\varrho^{2}]/z^{m+2}
\nonumber \\
\dotso \nonumber 
\end{eqnarray}

The first functions over the delocalized group coincide with the completely localized ones with the accuracy up to a multiplier:

\begin{equation}
{\cal A}^{(1,k)}_{1...k+1} = {\cal F}^{(k+1)}_{1...k+1}/z^{k+1}.
\label{eq:033}
\end{equation}

Equations (\ref{eq:019}),(\ref{eq:031}) - (\ref{eq:033}) show us that, evidently, functions ${\cal A}^{(m,k)}_{1...m+k}$ , in a certain sense, fill all the space  between complete and localized correlation functions.

The counterpart of (\ref{eq:025}) can be easily obtained from the definition (\ref{eq:031}):

\begin{equation}
\int\psi^V_{m+k} {\cal A}^{(m,k)}_{G,1...m+k}(\psi^V) d\bm{r}_{m+k} =  \frac{\partial{\cal A}^{(m,k-1)}_{G,1...m+k-1}(\psi^V)}{\partial z }  .
\label{eq:034}
\end{equation}

Hence follows the counterpart of (\ref{eq:026}):

\begin{eqnarray}
\int\left [ \prod_{i = l}^{m+k} \psi^V_i \right ] {\cal A}^{(m,k)}_{G,1...m+k}\mkern -8mu &(&\mkern -8mu\psi^V)d\bm{r}_{l}...d\bm{r}_{m+k} \label{eq:035} \\
&=&   \frac{\partial^{m+k-l+1}{\cal A}^{(m,l-m-1)}_{G,1...l-1}(\psi^V)}{\partial z^{m+k-l+1}},
\nonumber
\end{eqnarray}
where $l = m+1, m+2, ..., m+k$, and for (\ref{eq:027}) we write the analogous expression:
\begin{eqnarray}
\int\left [ \prod_{i = m+1}^{m+k} \psi^V_i \right ] {\cal A}^{(m,k)}_{G,1...m+k}(\psi^V\mkern -8mu&)&\mkern -8mu d\bm{r}_{m+1}...d\bm{r}_{m+k} \label{eq:036}\\
&=&  \frac{\partial^{k}{\cal A}^{(m,0)}_{G,1...m}(\psi^V)}{\partial z^{k}},
\nonumber
\end{eqnarray}
or, taking (\ref{eq:032}) into consideration yields in:  
\begin{eqnarray}
\int \left [ \prod_{i = m+1}^{m+k} \psi^V_i \right ]{\cal A}^{(m,k)}_{G,1...m+k}(\psi^V\mkern -7mu&)&\mkern -7mu d\bm{r}_{m+1}...d\bm{r}_{m+k} \label{eq:037} \\
&=&  \frac{\partial^{k}}{\partial z^{k}} \left [ \frac{{\varrho}^{(m)}_{G,1...m}(\psi^V)}{z^m} \right ].
\nonumber
\end{eqnarray} 

We shall also underline that in order for these relations to be true integration should be only carried over the coordinates of localized particles. The fact that the degrees of activity vanish from (\ref{eq:034}) - (\ref{eq:036}) in contrast to (\ref{eq:025}) - (\ref{eq:027}) gives us an insight into the purpose of the normalization.

The generating function for ${\cal F}^{(k)}_{1...k}$ is the logarithm, as we can see from (\ref{eq:011}), (\ref{eq:014}) and (\ref{eq:022}). The generating function for ${\cal A}^{(m,k)}_{1...m+k}$ - is the fraction of the following type:
\begin{eqnarray}
\frac{\displaystyle \sum_{k=0}^\infty \frac{(-1)^k}{k!} \int \left [ \prod_{i = 1}^{m+k} \psi^v_i \right ] \varrho^{(m+k)}_{1...m+k} d\bm{r}_1...d\bm{r}_{m+k}}{\displaystyle 1 + \sum_{k=1}^\infty \frac{(-1)^k}{k!}  \int \left [ \prod_{i = 1}^k \psi^v_i \right ] \varrho^{(k)}_{1...k} d\bm{r}_1 ...d\bm{r}_k}& & \label{eq:038}\\
= z^m\sum_{k=0}^\infty \frac{(-z)^k}{k!} \int \left [ \prod_{i = 1}^{m+k} \psi^v_i \right ] {\cal A}^{(m,k)}_{1...m+k} &d\bm{r}_1&...d\bm{r}_{m+k}.
\nonumber
\end{eqnarray}

As it can be easily obtained from (\ref{eq:038}) by multiplying the series this equation is ensured by the recurrence relation
\begin{eqnarray}
{\cal A}^{(m,k)}_{1...m+k} &=& {\cal A}^{(m+k,0)}_{1...m+k} \label{eq:039} \\
&-& \sum_{n=1}^k \sum_{perm} {\cal A}^{(n,0)}_{1...n}{\cal A}^{(m,k-n)}_{n+1...m+k} , \nonumber
\end{eqnarray}
where we employed the expression for ${\cal A}^{(n,0)}$ from (\ref{eq:032}), while the internal summing is performed over the permutations of localized particles only.

Relation (\ref{eq:039}) can be proved either by means of direct exhaustion of all decompositions in accordance with (\ref{eq:031}), or by the way of a repeated substitution of the expression for ${\cal A}^{(m,k)}_{1...m+k}$ in the right-hand side of (\ref{eq:039}).

Work \cite{percus1964} features another recurrence relation in regard to Ursell functions that, upon translating it to the language of ${\cal A}^{(m,k)}_{1...m+k}$, may be written as:
\begin{eqnarray}
{\cal A}^{(1,k)}_{1...k+1} &=&  {\cal A}^{(k+1,0)}_{1...k+1}\label{eq:040} \\
&-& \sum_{n=1}^{k} \binom{k}{n} \left [ {\cal A}^{(n,0)}_{1...n} {\cal A}^{(1,k-n)}_{n+1...k+1}\right ]_{perm}, \nonumber
\end{eqnarray}
where the brackets stand for the averaging over the permutations of particles.

As a matter of fact, it is the very same equation (\ref{eq:039}) at $m=1$ but written in a symmetric form. The question about the necessity of the symmetrization of all recurrence relations for ${\cal A}^{(m,k)}_{1...m+k}$, which we will further face, falls beyond the scope of the present work; note, though, that in the given case there is no need in symmetrization as Eq. (\ref{eq:039}) is strictly satisfied in the assymetrical form. Nevertheless, (\ref{eq:039}) can also be written in the symmetric form, but this will require the averaging of the left-hand side of the equation and not that over the sum permutations only.

The satisfaction of these and many other recurrence relations for ${\cal A}^{(m,k)}_{1...m+k}$ is crucial for the mathematical support of the approach being presented. As it will become clear further in the course of our discussion, they are also true for the corresponding factors and give us the opportunity to operate with the expressions from the OSE.

\subsection{\label{subsec:03d}Factors}

Replacing ${\varrho}^{(k)}_{1...k}$ by $z^k\exp(-\beta U^{k}_{1...k})$ in the defining equations (\ref{eq:019}),  (\ref{eq:031}) brings for the following factors:
Boltzmann's:  

\begin{equation}
\exp(-\beta U^{k}_{1...k}) \leftrightarrow {\varrho}^{(k)}_{1...k}/z^k,
\label{eq:041}
\end{equation}
Ursell's:  

\begin{equation}
{\cal U}^{(k)}_{1...k} \leftrightarrow {\cal F}^{(k)}_{1...k}/z^k
\label{eq:042}
\end{equation} 
and partial localization factors:  

\begin{equation}
{\cal B}^{(m,k)}_{1...m+k} \leftrightarrow {\cal A}^{(m,k)}_{1...m+k}
\label{eq:043}.
\end{equation} 
 
At the same time

\begin{equation}
1 \leftrightarrow \varrho/z.
\label{eq:044}
\end{equation}

In other words, let us define ${\cal B}^{(m,k)}_{1...m+k}$ by an equation:

\begin{eqnarray}
{\cal B}^{(m,k)}_{1...m+k} &=& \sum_{\{\bm{n}\}}(-1)^{l-1}(l-1)!\label{eq:045}\\
&\times&\prod_{\alpha = 1}^l \exp(-\beta U^{k_\alpha+(m-1)\delta_{\alpha\gamma}}(\{\bm{n}_\alpha\}))  \nonumber \\
&& 1  < ~ k_\alpha \leq k+1, ~ \alpha \neq \gamma ~ or ~ m=1  \nonumber \\
&& 1  \leq ~ k_\gamma \leq k+1, m > 1; \sum_{\alpha = 1}^l k_\alpha \leq k+1,\nonumber 
\end{eqnarray}
whose notations are analogous to those of (\ref{eq:031}). The differences are observed due to the fact that one-particle groups do not have any contribution to the products in this case.
 
As factors are primary values independent of the ensemble type or window functions, we omit these notations for them.

The generating function for ${\cal B}^{(m,k)}_{1...m+k}$, as follows from (\ref{eq:038}), (\ref{eq:013}), (\ref{eq:041}) is a complete correlation function of the GCE-type

\begin{widetext}
\begin{equation}
\frac{\varrho^{(m)}_{G,1...m}}{z^m} = \frac{\displaystyle \sum_{k=0}^\infty \frac{z^k}{k!} \int \left [ \prod_{i = m+1}^{m+k} \psi^V_i \right ] e^{-\beta U^{m+k}_{1...m+k}} d\bm{r}_{m+1}...d\bm{r}_{m+k}}{\displaystyle 1 + \sum_{k=1}^\infty \frac{z^k}{k!}  \int \left [ \prod_{i = 1}^k \psi^V_i \right ] e^{-\beta U^{k}_{1...k}} d\bm{r}_1 ...d\bm{r}_k} = {\cal B}^{(m,0)}_{1...m} + \sum_{k=1}^\infty \frac{z^k}{k!} \int \left [ \prod_{i = m+1}^{m+k} \psi^V_i \right ] {\cal B}^{(m,k)}_{1...m+k} d\bm{r}_{m+1}...d\bm{r}_{m+k}.
\label{eq:046}
\end{equation}
\end{widetext}

Ursell factors, also known as cluster functions, are involved in a well-known expansion of pressure in powers of activity \cite[p.129]{hillstatmeh1987}, \cite[p.232]{landaulifshitz1985}. 

Boltzmann factors are also well known and make part of configuration integrals (\ref{eq:003}). 

The factors of partial localization, however, have not yet been investigated - at least, as far as we are concerned.
 
For instance: 
\begin{eqnarray}
{\cal U}^{(1)}_{1}\mkern 9mu &=& 1 \label{eq:047} \\
{\cal U}^{(2)}_{1,2} \mkern 8mu &=& \exp(-\beta U^{2}_{1,2}) - 1 \nonumber \\
{\cal U}^{(3)}_{1,2,3} &=& \exp(-\beta U^{3}_{1,2,3}) - \exp(-\beta U^{2}_{1,2}) \nonumber \\
 &-& \exp(-\beta U^{2}_{1,3}) - \exp(-\beta U^{2}_{2,3}) + 2 \nonumber \\
\dotso \nonumber
\end{eqnarray}

The first ${\cal B}^{(m,k)}_{1...m+k}$'s over the localized group:
\begin{eqnarray}
{\cal B}^{(m,0)}_{1...m} \mkern 12mu &=& \exp(-\beta U^{m}_{1...m})\label{eq:048} \\
{\cal B}^{(m,1)}_{1...m+1} &=& \exp(-\beta U^{m+1}_{1...m+1}) - \exp(-\beta U^{m}_{1...m})\nonumber \\
{\cal B}^{(m,2)}_{1...m+2} & = & \exp(-\beta U^{m+2}_{1...m+2}) -  \exp(-\beta U^{m+1}_{1...m+1})\nonumber \\
 &-& \exp(-\beta U^{m+1}_{1...m,m+2}) - \exp(-\beta U^{m}_{1...m}) \nonumber \\
 &\times& \exp(-\beta U^{2}_{m+1,m+2})  + 2\exp(-\beta U^{m}_{1...m}) \nonumber \\
\dotso \nonumber
\end{eqnarray} 
and over the delocalized one:
\begin{equation}
{\cal B}^{(1,k-1)}_{1...k} = {\cal U}^{(k)}_{1...k}
\label{eq:049},
\end{equation}
including 
\begin{equation}
{\cal B}^{(1,0)}_{1} = 1
\label{eq:050}.
\end{equation}
where we showed the form for a uniform environment.

As it will be later shown, the factors constitute the limits of the corresponding correlation functions that are normalized by the required degree $z$ with activity tending to zero. In their turn, ${\cal A}^{(m,k)}_{1...m+k}$ do not require normalization in order for this to be true. 

In the same manner to ${\cal A}^{(m,k)}_{1...m+k}$, factors ${\cal B}^{(m,k)}_{1...m+k}$ generalize Ursell and Boltzmann factors incorporating them as extreme cases.  

Evidently, factors maintain the algebraic properties of correlation functions and thus the recurrence relations identical to (\ref{eq:039}) and (\ref{eq:040}) hold true for them as well:

\begin{eqnarray}
{\cal B}^{(m,k)}_{1...m+k} &=& {\cal B}^{(m+k,0)}_{1...m+k} \label{eq:051} \\
&-& \sum_{n=1}^k \sum_{perm} {\cal B}^{(n,0)}_{1...n}{\cal B}^{(m,k-n)}_{n+1...m+k} , \nonumber
\end{eqnarray}
true when $m \geq 1$, $k \geq 1$ and
\begin{eqnarray}
{\cal B}^{(1,m)}_{1...m+1} &=& {\cal B}^{(m+1,0)}_{1...m+1} \label{eq:052} \\
&-& \sum_{n=1}^{m} \binom{m}{n} \left [ {\cal B}^{(n,0)}_{1...n} {\cal B}^{(1,m-n)}_{n+1...m+1}\right ]_{perm}, \nonumber
\end{eqnarray}
where $m \geq 1$.

\def\dosimm{\stackrel{sim}{=}}\def\convf{\hbox{\space \raise-2mm\hbox{$\textstyle      \bigotimes \atop \scriptstyle \omega$} \space}}

As it has been already said, the question about the fulfillment of recurrence relations in the asymmetrical form of type (\ref{eq:051}) extends beyond the frame of this work. It is not improbable that all of the further given relations have this form but we will be using the symmetric one. In a number of cases we will make use of the symbol $\dosimm$; at that we will imply that the right-hand and the left-hand sides of the equation underwent the symmetrization over those indexes where it is necessary. In this case, symmetrization will not only mean the averaging over the permutations but also aligning the indexes to the natural scale if it is required. This operation appears as a consequence to property (\ref{eq:050}).

So, apart from (\ref{eq:051}), (\ref{eq:052}) the following equations are fulfilled: 
\begin{eqnarray}
{\cal B}^{(m,k)}_{1...m+k} &\dosimm& {\cal B}^{(m-1,k+1)}_{1...m+k} \label{eq:053} \\
&+& \sum_{n=0}^k \binom{k}{n}  {\cal B}^{(m-1,k-n)}_{1...m+k-n-1} {\cal B}^{(1,n)}_{m+k-n...m+k}, \nonumber
\end{eqnarray}
at $k \geq 0$ and $m \geq 2$, as well as 
\begin{eqnarray}
{\cal B}^{(1,k)}_{1...k+1} &=&   \sum_{m=1}^{k} \binom{k}{m} (-1)^m \label{eq:054} \\
&\times& \left [ {\cal B}^{(m,k-m)}_{1...k} - {\cal B}^{(m+1,k-m)}_{1...k+1}\right ]_{perm}, \nonumber
\end{eqnarray}
where $k \geq 1$; and we also witness the satisfaction of two classes of recurrence relations that have two free indexes with the former class containing the low-value indexes of delocalized particles: 
\begin{eqnarray}
\sum_{m=0}^{k}&(-1)^{m}& \binom{k}{m} {\cal B}^{(m+1,t+k-m)}_{1...t+k+1} ~ \dosimm ~ \sum_{l=1}^{k} \binom{k}{l}  \label{eq:055}\\
\times &(-1)^{l}& \sum_{n=0}^{t} \binom{t}{n}  {\cal B}^{(l,k+n-l)}_{1...k+n} {\cal B}^{(1,t-n)}_{k+n+1...k+t+1}, \nonumber
\end{eqnarray}
where $k \geq 1$, $t\geq 0$,  and for $k = 1$ it reduces to (\ref{eq:053}) with $m=2$, while for $t = 0$ it amounts to (\ref{eq:054}). The latter class, au contraire, contains those of localized particles:
\begin{eqnarray}
\sum_{m=0}^{k}&(-1)^{m}& \binom{k}{m} {\cal B}^{(m+t+1,k-m)}_{1...t+k+1}~ \dosimm ~\sum_{l=0}^{k} \binom{k}{l}  \label{eq:056}\\
\times &(-1)^{l}& \sum_{n=1}^{t} \binom{t}{n}  {\cal B}^{(l+n,k-l)}_{1...k+n} {\cal B}^{(1,t-n)}_{k+n+1...k+t+1} \nonumber \\
&+& \sum_{s=1}^{k} (-1)^{s} \binom{k}{s} {\cal B}^{(s,k-s)}_{1...k} {\cal B}^{(1,t)}_{k+1...k+t+1}, \nonumber
\end{eqnarray}
where $k \geq 1$, $t\geq 1$, and for $k = 1$ it reduces to (\ref{eq:052}). 

Equations (\ref{eq:053}) - (\ref{eq:056}) will be proved in the section  \ref{subsec:05d}.

All of them ensure the fulfillment of certain physical relations that will be defined later.

\section{\label{sec:04}The Formulation of the OSE}

\subsection{\label{subsec:04a}The Partition Function of the OSE}

Now we are able to compute $p^v_0$ and consequently define the partition function of the ensemble. This function will further be referred to as $\Upsilon_v$, where $v$ stands for the volume of the system: 
\begin{equation}
\Upsilon_v =  \frac{1}{p^v_0}.
\label{eq:057}
\end{equation}

Let us employ the expansion of functions ${\cal F}^{(k)}_{G,1...k}(\psi^V)$ that is given in \cite{LebowitzPenrose1968}. Writing it in the language of $\psi^V$ gives:
\begin{eqnarray}
{\cal F}^{(k)}_{G,1...k}(\psi^V)&=& z^k{\cal  U}^{(k)}_{1...k} + z^k\sum_{n=1}^\infty \frac{z^n}{n!} \label{eq:058}  \\
&\times& \int \left [ \prod_{l = k+1}^{k+n} \psi^V_l \right ] {\cal  U}^{(k+n)}_{1...k+n} d\bm{r}_{k+1}...d\bm{r}_{k+n}.    \nonumber
\end{eqnarray}

By substituting (\ref{eq:058}) in (\ref{eq:014}) and changing the order of summing we have:
\begin{eqnarray}
\ln{p^v_{G,0}} &=&  \sum_{t=1}^\infty \frac{z^t}{t!}\int{\cal U}^{(t)}_{1...t} \label{eq:059} \\
&\times & \left ( \sum_{k=1}^t \binom{t}{k} (-1)^k \prod_{i = 1}^k \prod_{j = k+1}^{t} \psi^v_i  \psi^V_j  \right ) d\bm{r}_1...d\bm{r}_t. \nonumber
\end{eqnarray}

Recasting (\ref{eq:059}) with the aid of the binomial formula, and utilizing the symmetry of ${\cal  U}^{(t)}_{1...t}$ with respect to particle permutations, results in the expression:
\begin{eqnarray}
\ln{p^v_{G,0}} &=& - \sum_{t=1}^\infty \frac{z^t}{t!} \int \left [ \prod_{i = 1}^{t} \psi^V_i \right ]  {\cal U}^{(t)}_{1...t} d\bm{r}_1...d\bm{r}_t \label{eq:060} \\
&+& \sum_{t=1}^\infty \frac{z^t}{t!} \int \left [ \prod_{i = 1}^{t} (\psi^V_i - \psi^v_i) \right ] {\cal U}^{(t)}_{1...t} d\bm{r}_1...d\bm{r}_t,\nonumber
\end{eqnarray}
or,
\begin{equation}
\ln{p^v_{G,0}} = -\ln{\Xi_V} + \ln{\Xi_{V-v}}.
\label{eq:061}  
\end{equation}

Indeed, both terms in the right-hand side of (\ref{eq:060}) constitute the logarithms of the partition functions of the GCE, since, taking the logarithm of Eq. (\ref{eq:010}), and taking (\ref{eq:003}) into account as well as relation (\ref{eq:022}) yields:
\begin{equation}
\ln{\Xi_V} =  \sum_{t=1}^\infty \frac{z^t}{t!} \int \left [ \prod_{i = 1}^{t} \psi^V_i \right ]  {\cal U}^{(t)}_{1...t} d\bm{r}_1...d\bm{r}_t.
\label{eq:062}  
\end{equation}

Volume terms of Eq. (\ref{eq:062}) give us the expansion for pressure \cite[p.135]{hillstatmeh1987} but we will extract them a little bit later.
The formulation of the OSE is concerned with the following limit: 
\begin{equation}
\ln{\Upsilon_v} =  \lim_{V\rightarrow \infty} \left ( \ln{\Xi_V} - \ln{\Xi_{V-v}} \right ),
\label{eq:063}  
\end{equation}
while the intensive parameters of the environment and the parameters of volume $v$ are maintained, or course. In order to proceed to the limit we unite the integrals from the right-hand side of (\ref{eq:060}) and assume that $\psi^V_i = 1$, for all $i$:
\begin{equation}
\ln{\Upsilon_v} = \sum_{t=1}^\infty \frac{z^t}{t!} \int \left [1 - \prod_{i = 1}^{t} (1 - \psi^v_i) \right ] {\cal U}^{(t)}_{1...t} d\bm{r}_1...d\bm{r}_t.
\label{eq:064} 
\end{equation}

Thus, we obtained one of the expressions for the partition function of the OSE in a uniform environment:
\begin{equation}
\Upsilon_v = \exp {\left \{ \sum_{t=1}^\infty \frac{z^t}{t!} \int \left [1 - \prod_{i = 1}^{t} (1 - \psi^v_i) \right ] {\cal U}^{(t)}_{1...t} d\bm{r}_1...d\bm{r}_t \right \} }.
\label{eq:065} 
\end{equation}

Now let us split it into volume and surface parts by reconstituting (\ref{eq:064}) in the following way:
\begin{eqnarray}
\ln{\Upsilon_v} &=& \sum_{t=1}^\infty \frac{z^t}{t!} \int  \psi^v_1 {\cal U}^{(t)}_{1...t} d\bm{r}_1...d\bm{r}_t \label{eq:066} \\
&+& \sum_{t=2}^\infty \frac{z^t}{t!} \int \left [1-\psi^v_1 - \prod_{i = 1}^{t} (1 - \psi^v_i) \right ] {\cal U}^{(t)}_{1...t} d\bm{r}_1...d\bm{r}_t.\nonumber
\end{eqnarray}

The first term, being the volume one, brings forth the well-known expansion for pressure:
\begin{equation}
Pv = k_BT\sum_{t=1}^\infty \frac{z^t}{t!} \int \psi^v_1{\cal U}^{(t)}_{1...t} d\bm{r}_1...d\bm{r}_t 
\label{eq:067}  
\end{equation}
or:
\begin{equation}
P(z,T) =zk_BT  + k_BT \sum_{k=2}^\infty \frac{z^k}{k!} \int {\cal U}^{(k)}_{1...k} d\bm{r}_2...d\bm{r}_k,
\label{eq:068} 
\end{equation}
it has, however, the counterpart for surface tension written below.
Let us factor out $1 - \psi^v_1$ from the second term and replace it by $\chi^v_1$, where  
\begin{equation}
\chi^v_i = 1 - \psi^v_i = 
	 \left\{ 
			\begin{array}{ll} 
         0 & (\bm{r}_i \in v)\\   
         1 & (\bm{r}_i \notin v).
     	\end{array}  
		\right.
		\label{eq:069}
\end{equation}

If function $\psi^v_i$ localizes the $i$-th particle inside volume $v$, then $\chi^v_i$ does the same outside its limits. We shall further have:
\begin{eqnarray}
\ln{\Upsilon_v} &=& Pv/k_BT \label{eq:070} \\
&+& \sum_{t=2}^\infty \frac{z^t}{t!} \int \chi^v_1 \left [1 - \prod_{i = 2}^{t} (1 - \psi^v_i) \right ] {\cal U}^{(t)}_{1...t} d\bm{r}_1...d\bm{r}_t.\nonumber
\end{eqnarray}

The second term in the right-hand side of (\ref{eq:070}) is proportional to the surface area. Indeed, its structure prescribes at least two particles to lie at the different sides of the boundary. If we expand the product in (\ref{eq:070}), then all terms will be written:
\begin{equation}
\int \chi^v_1 \left [ \prod_{i = 2}^{j} \psi^v_i \right ] {\cal U}^{(t)}_{1...t} d\bm{r}_1...d\bm{r}_t,
\label{eq:071}  
\end{equation}
where $2 \leq j \leq t $. When the first particle is fixed, the integrals of type (\ref{eq:071}) are defined by the local area in the vicinity of this particle due to the local character of Ursell factors. As a consequence, $\int ... d\bm{r}_2...d\bm{r}_t$ do not depend on displacements of the first particle along the boundary of the system. However, when the first particle moves away from the boundary they quickly die away because of the fixing factors $\psi^v_i$ and the local character of Ursell factors. Performing the integration alongside the surface and taking the area out of the summation sign yields:
\begin{equation}
\Upsilon_v = \exp{\beta [ vP(z,T)  + a\sigma (z,T) ]},
\label{eq:072}
\end{equation}
where $a$ - is the area delimiting the system, $vP(z,T)$ is given by Eq. (\ref{eq:067}), $a\sigma (z,T)$ shall be written
\begin{eqnarray}
a\sigma (z,T) &=& k_BT\sum_{t=2}^\infty \frac{z^t}{t!} \label{eq:073} \\
& \times &  \int \chi^v_1 \left [1 - \prod_{i = 2}^{t} (1 - \psi^v_i) \right ] {\cal U}^{(t)}_{1...t} d\bm{r}_1 ...d\bm{r}_t \nonumber
\end{eqnarray}
and for $\sigma (z,T)$ we have the following expression:
\begin{eqnarray}
\sigma (z,T) &=& k_BT\sum_{t=2}^\infty \frac{z^t}{t!} \label{eq:074} \\
&\times &  \int \chi^v_1 \left [1 - \prod_{i = 2}^{t} (1 - \psi^v_i) \right ] {\cal U}^{(t)}_{1...t} dx_1 d\bm{r}_2 ...d\bm{r}_t,\nonumber
\end{eqnarray}
where $x_1$ - is the coordinate perpendicular to the surface given by the boundary of $\psi^v_i$. The direction of axis $x_1$ is chosen in the way to satisfy $dx_1 > 0$.

The first terms of expression (\ref{eq:074}) were obtained in \cite{Bellemans1962} by means of the diagram technique, but building up large number terms in the series with the aid of this approach seems to be extremely difficult.

For the cavity formation probability we shall write:
\begin{equation}
p^v_0 = \exp{- \beta [vP(z,T) + a \sigma (z,T)]}.
\label{eq:075}
\end{equation}

We will discuss this result from the thermodynamic point of view later on.

By making use of $\chi^v$, the expression (\ref{eq:064}) takes the following form:
\begin{equation}
\ln{\Upsilon_v} = \sum_{t=1}^\infty \frac{z^t}{t!} \int \left [1 - \prod_{i = 1}^{t} \chi^v_i \right ] {\cal U}^{(t)}_{1...t} d\bm{r}_1...d\bm{r}_t,
\label{eq:076} 
\end{equation}
which is interesting to compare with the expression for the logarithm of the partition function of the GCE (\ref{eq:062}).

\subsection{\label{subsec:04b}The Distribution of the OSE}

We can now calculate the common term of the distribution of the OSE. So, consider the probability to find $m$ particles in volume $v$ for a uniform system - $p^v_m$. The outline of operations here will correspond to the reckoning of the zero term given above.
\begin{eqnarray}
p^v_{C,m}& = &\binom{N}{m}\frac{1}{Z_N}\int\exp(-\beta U^N_{1...N})\label{eq:077}  \\
 &\times&  \left [\prod_{i = 1}^m\prod_{j = m+1}^N\psi^v_i (\psi^V_j-\psi^v_j) \right ]  d\bm{r}_1...d\bm{r}_N. \nonumber
\end{eqnarray}

The binomial coefficient appears in (\ref{eq:077}) as a consequence of us considering only specific sets of particles at this stage. By calculating in the same manner to (\ref{eq:004}), (\ref{eq:007}) we have:
\begin{eqnarray}
p^v_{C,m} &=& \frac{1}{m!} \sum_{k=0}^{N-m} \frac{(-1)^k}{k!}  \label{eq:078} \\
&\times & \int \left [ \prod_{i = 1}^{m+k} \psi^v_i \right ] \varrho^{(m+k)}_{C,1...m+k} d\bm{r}_1...d\bm{r}_{m+k}. \nonumber
\end{eqnarray}

Shifting to the GCE and changing the order of summing yields the expression analogous to (\ref{eq:011}):
\begin{eqnarray}
p^v_{G,m} &=& \frac{1}{m!} \sum_{k=0}^\infty \frac{(-1)^k}{k!}  \label{eq:079} \\
&\times & \int \left [ \prod_{i = 1}^{m+k} \psi^v_i \right ] \varrho^{(m+k)}_{G,1...m+k}(\psi^V) d\bm{r}_1...d\bm{r}_{m+k}.\nonumber
\end{eqnarray}

We further divide the series (\ref{eq:079}) by (\ref{eq:011}) and thus obtain the relation:
\begin{eqnarray}
\frac{m!p^v_{G,m}}{z^m p^v_{G,0}} &=& \sum_{k=0}^\infty \frac{(-z)^k}{k!}  \label{eq:080} \\
&\times & \int \left [ \prod_{i = 1}^{m+k} \psi^v_i \right ] {\cal A}^{(m,k)}_{G,1...m+k} (\psi^V) d\bm{r}_1...d\bm{r}_{m+k},
\nonumber
\end{eqnarray}
where $m = 1,2,..$. We resorted here to the fractionary generating function (\ref{eq:038}) for the GCE-type correlation functions.

Now we shall derive the expression analogous to (\ref{eq:058}), which generalizes it for ${\cal A}^{(m,k)}_{G,1...m+k}(\psi^V)$.

In order to do this we shall expand this function in a Taylor series in powers of activity
\begin{eqnarray}
{\cal A}^{(m,k)}_{G,1...m+k}(z,\psi^V) &=& {\cal A}^{(m,k)}_{G,1...m+k}(0,\psi^V) \label{eq:081} \\
&+& \sum_{n=1}^\infty \frac{z^n}{n!}  \left[\frac{\partial^{n}}{\partial z^{n}} {\cal A}^{(m,k)}_{G,1...m+k}(z,\psi^V) \right]_{z=0} 
\nonumber
\end{eqnarray}
and taking (\ref{eq:035}) into account yields:
\begin{eqnarray}
&&\mkern -36mu{\cal A}^{(m,k)}_{G,1...m+k}(z,\psi^V) = {\cal A}^{(m,k)}_{G,1...m+k}(0,\psi^V) \label{eq:082}  \\
&&+ \sum_{n=1}^\infty \frac{z^n}{n!} \int \left [ \prod_{i = m+k+1}^{m+k+n} \psi^V_i \right ]\nonumber\\
&&\mkern +36mu\times{\cal A}^{(m,k+n)}_{G,1...m+k+n}(0,\psi^V) d\bm{r}_{m+k+1}...d\bm{r}_{m+k+n}. \nonumber
\end{eqnarray}

Then, the definition (\ref{eq:013}), at $z \rightarrow 0$, leads to 
\begin{equation}
\varrho^{(k)}_{1...k} \rightarrow z^k\exp(-\beta U^{k}_{1...k}) ,
\label{eq:083}
\end{equation}
\cite[p.410]{hillstatmeh1987}. As decompositions (\ref{eq:031}) always maintain the total rank $\varrho^{(k)}_{1...k}$ we may conclude that 
\begin{equation}
{\cal A}^{(m,k)}_{1...m+k}(0) =  {\cal B}^{(m,k)}_{1...m+k},
\label{eq:084}
\end{equation}  
and thus we obtain the relation we initially sought for:
\begin{eqnarray}
&&\mkern -12mu{\cal A}^{(m,k)}_{G,1...m+k}(z,\psi^V) = {\cal B}^{(m,k)}_{1...m+k} + \sum_{n=1}^\infty \frac{z^n}{n!}\label{eq:085}  \\
&&\mkern +12mu\times \int \left [ \prod_{i = m+k+1}^{m+k+n} \psi^V_i \right ]{\cal B}^{(m,k+n)}_{1...m+k+n} d\bm{r}_{m+k+1}...d\bm{r}_{m+k+n}. \nonumber
\end{eqnarray}
 
At $k>0$, $m=1$, when taking (\ref{eq:033}) and (\ref{eq:049}) into consideration, we obtain (\ref{eq:058}); at $k=0$, $m>1$ and by taking (\ref{eq:032}) into account it yields the expansion for $\varrho^{(k)}_{G,1...k}(\psi^V)$, 
\begin{eqnarray}
\varrho^{(m)}_{G,1...m}(z,\psi^V\mkern -7mu&)&\mkern -7mu = z^m {\cal B}^{(m,0)}_{1...m} + z^m\sum_{n=1}^\infty \frac{z^n}{n!}\label{eq:086}  \\
\mkern -7mu&\times&\mkern -7mu \int\left [ \prod_{i = m+1}^{m+n} \psi^V_i \right ] {\cal B}^{(m,n)}_{1...m+n} d\bm{r}_{m+1}...d\bm{r}_{m+n}, \nonumber
\end{eqnarray}
which exactly agrees with (\ref{eq:046}).

Finally, at $k=0$, $m=1$ the well-known expansion for the density of a uniform system follows from (\ref{eq:085}) if we take (\ref{eq:049}) and (\ref{eq:050}) into consideration: 

\begin{equation}
\varrho_G(z,\psi^V) = z + z\sum_{n=1}^\infty \frac{z^n}{n!} \int \left [ \prod_{i = 2}^{n+1} \psi^V_i \right ]{\cal U}^{(n+1)}_{1...n+1} d\bm{r}_{2}...d\bm{r}_{n+1}.
\label{eq:087}
\end{equation}(Evidently, (\ref{eq:087}) may also follow from (\ref{eq:058}) at $k=1$). In this relation the factors $\psi^V_i$ are often neglected or $\varrho_G(\psi^V)$ is substituted for $\varrho$, or both.

Now we can substitute (\ref{eq:085}) in (\ref{eq:080}), and, by following the steps similar to proceeding from (\ref{eq:014}) via (\ref{eq:059}) to (\ref{eq:060}) we obtain: 
\begin{eqnarray}
&&p^v_{G,m} = \frac{z^m}{m! \Upsilon_v}\sum_{t=0}^\infty \frac{z^t}{t!} \label{eq:088} \\
&&\times \int \left [ \prod_{i = 1}^m \prod_{j = m+1}^{m+t}\psi^v_i(\psi^V_j-\psi^v_j) \right ] {\cal B}^{(m,t)}_{1...m+t} d\bm{r}_1...d\bm{r}_{m+t}, \nonumber
\end{eqnarray}
where $m = 1,2...$. The passage to the limit  
\begin{eqnarray}
&& V \rightarrow \infty, \label{eq:089} \\
&& v = const\nonumber, \\
&& ip = const\nonumber,
\end{eqnarray}
where $ip$ are the intensive parameters of the environment, yields
\begin{eqnarray}
p^v_m &=& \frac{z^m}{m! \Upsilon_v}\sum_{t=0}^\infty \frac{z^t}{t!} \label{eq:090} \\
&\times& \int \left [ \prod_{i = 1}^m \prod_{j = m+1}^{m+t}\psi^v_i\chi^v_j \right ] {\cal B}^{(m,t)}_{1...m+t} d\bm{r}_1...d\bm{r}_{m+t}. \nonumber
\end{eqnarray}

This expression solves the posed problem by defining the general term of the distribution of the OSE.

It is easy to see that the first term of the series (\ref{eq:090}) corresponds to the distribution of the GCE (\ref{eq:008}) due to (\ref{eq:048}) or because of (\ref{eq:050}) at $m=1$ with the accuracy up to normalizing factors - partition functions.
 	
The structure of the series (\ref{eq:090}) seems to be of interest. The location of the delocalized group of $m$ particles is set by window functions $\psi^v_i$ inside the volume of the system. The location of the localized group of $t$ particles is, in its turn, defined outside it by functions $\chi^v_j$, being, however, closely bound with the volume due to functions ${\cal B}^{(m,t)}_{1...m+t}$. The summing is performed over all clusters that are bigger than $m$. It may be stated, that all volume-related properties of the expansion are assigned by the first $m$ particles - ordinary, delocalized ones, whereas surface properties are predetermined by running $t$ particles, that are localized.

As the differences between the GCE and the OSE are related to surface terms, all the elements of the series (\ref{eq:090}), starting from the second one, contain the volume as well as the surface of the system in different degrees.

\section{\label{sec:05}Certain Properties of the OSE}

Making calculations with the OSE is somewhat more difficult than with the GCE. Every single operation with this distribution is ensured by a certain class of recurrence relations for ${\cal B}^{(m,k)}_{1...m+k}$. For brevity sake, we say that an operation yields a recurrence relation or class. On the other hand, the arising difficulties can be overcome in a number of cases by means of using various generating functions. Consider some of these operations.

\subsection{\label{subsec:05a}The Condition of Normalization}

$\sum p^v_m$ over all $m$'s shall be equal to 1, and thus we obtain the equation:
\begin{eqnarray}
\Upsilon_v &=&1 + \sum_{m=1}^\infty \frac{z^m}{m!}\sum_{t=0}^\infty \frac{z^t}{t!} \label{eq:091} \\
&\times& \int \left [ \prod_{i = 1}^m \prod_{j = m+1}^{m+t}\psi^v_i\chi^v_j \right ] {\cal B}^{(m,t)}_{1...m+t} d\bm{r}_1...d\bm{r}_{m+t}. \nonumber
\end{eqnarray}

This is yet another form for the partition function of the OSE. It is difficult to see here expression (\ref{eq:065})! 
In order to prove these formulas identical we may take their logarithmic derivatives. Then we have:
\begin{equation}
\frac{1}{\Upsilon_v}\frac{\partial\Upsilon_v}{\partial z} = \frac{\partial \beta(Pv+\sigma a)}{\partial z},
\label{eq:092}
\end{equation} 
or
\begin{equation}
\frac{\partial\Upsilon_v}{\partial z} = \Upsilon_v \frac{\partial \beta(Pv+\sigma a)}{\partial z},
\label{eq:093}
\end{equation} 
moreover, $\Upsilon_v$ in (\ref{eq:093}) shall be defined by expression (\ref{eq:091}), while $P v$ and $\sigma a$ - by (\ref{eq:067}) and (\ref{eq:073}) correspondingly, where ${\cal U}^{(t)}_{1...t}$ are substituted by ${\cal B}^{(1,t-1)}_{1...t}$ according to (\ref{eq:049}):
\begin{equation}
vP(z,T) = k_BT\sum_{t=1}^\infty \frac{z^t}{t!} \int \psi^v_1{\cal B}^{(1,t-1)}_{1...t} d\bm{r}_1...d\bm{r}_t 
\label{eq:094}  
\end{equation}
\begin{eqnarray}
a\sigma (z,T) &=& k_BT\sum_{t=2}^\infty \frac{z^t}{t!} \label{eq:095} \\
&+&  \int \chi^v_1 \left [1 - \prod_{i = 2}^{t} (1 - \psi^v_i) \right ] {\cal B}^{(1,t-1)}_{1...t} d\bm{r}_1 ...d\bm{r}_t.\nonumber
\end{eqnarray}

Let us change the order of summing in the expressions to facilitate the calculations. By the example of (\ref{eq:091}) we write:
\begin{eqnarray}
\Upsilon_v &=&1 + \sum_{s=1}^\infty \frac{z^s}{s!}\int\sum_{m=1}^s \binom{s}{m} \label{eq:096} \\
&\times&  \left [ \prod_{i = 1}^m \prod_{j = m+1}^{s}\psi^v_i\chi^v_j \right ] {\cal B}^{(m,s-m)}_{1...s} d\bm{r}_1...d\bm{r}_{s}. \nonumber
\end{eqnarray}

Differentiating and multiplying the series as well as equating the expressions at different degrees of $z$ and products $\psi^v_i$ yields recurrence relations (\ref{eq:053}), (\ref{eq:054}) and the class of relations (\ref{eq:056}).

By applying ordinary thermodynamic formulas to (\ref{eq:092}) we may write:
\begin{equation}
\frac{\partial\ln{\Upsilon_v}}{\partial \ln{z}} = m_b - m_s,
\label{eq:097}
\end{equation} 
where
\begin{equation}
m_b = v\frac{\partial P}{\partial \mu}
\label{eq:098}
\end{equation}  
- is the number of bulk particles, whereas
\begin{equation}
m_s = - a\frac{\partial \sigma}{\partial \mu}
\label{eq:099}
\end{equation}
- is that of the surface ones.

We will further compare (\ref{eq:097}) with a similar relation for the GCE.

\subsection{\label{subsec:05b}Mean Particle Number} 

The next operation to be considered is calculating $\overline{m}$. It yields the class of recurrence relations (\ref{eq:055}). We will proceed from the natural assumption that
\begin{equation}
\overline{m} = m_b.
\label{eq:100}
\end{equation}

The proof can be given by taking the following steps. According to (\ref{eq:090}) $\overline{m}$ takes the form:

\begin{eqnarray}
\overline{m} &=& \frac{1}{\Upsilon_v}\sum_{m=1}^\infty\frac{z^m}{(m-1)!} \sum_{t=0}^\infty \frac{z^t}{t!} \label{eq:101} \\
&\times& \int \left [ \prod_{i = 1}^m \prod_{j = m+1}^{m+t}\psi^v_i\chi^v_j \right ] {\cal B}^{(m,t)}_{1...m+t} d\bm{r}_1...d\bm{r}_{m+t}. \nonumber
\end{eqnarray}

Let us change the order of summing:
\begin{eqnarray}
\overline{m} &=& \frac{1}{\Upsilon_v}\sum_{s=1}^\infty\frac{z^s}{(s-1)!}\int \sum_{m=1}^s \binom{s-1}{m-1} \label{eq:102} \\
&&\times  \left [ \prod_{i = 1}^m \prod_{j = m+1}^{s}\psi^v_i\chi^v_j \right ] {\cal B}^{(m,s-m)}_{1...s} d\bm{r}_1...d\bm{r}_{s}. \nonumber
\end{eqnarray}
  
We shall equate this expression to (\ref{eq:098}), where we need to employ (\ref{eq:094}) for $P$. By multiplying the series and equating the expressions at different degrees of $z$ and products $\psi^v_i$ we obtain the class of recurrence relations (\ref{eq:055}).

Equation (\ref{eq:101}) yields a new expression for $m_s$. With substituting (\ref{eq:100}) in (\ref{eq:097}), we have:
\begin{equation}
\frac{\partial\ln{\Upsilon_v}}{\partial \ln{z}} = \overline{m} - m_s.
\label{eq:103}
\end{equation}

Further on, by using $\Upsilon_v$ in the form of (\ref{eq:091}) and $\overline{m}$ in the form of (\ref{eq:101}) we can obtain from (\ref{eq:103}) the following expressions:
\begin{eqnarray}
m_s &=& - \frac{1}{\Upsilon_v}\sum_{m=1}^\infty\frac{z^m}{m!} \sum_{t=1}^\infty \frac{z^t}{(t-1)!} \label{eq:104} \\
&\times& \int \left [ \prod_{i = 1}^m \prod_{j = m+1}^{m+t}\psi^v_i\chi^v_j \right ] {\cal B}^{(m,t)}_{1...m+t} d\bm{r}_1...d\bm{r}_{m+t}, \nonumber
\end{eqnarray}
or
\begin{eqnarray}
m_s &=& - \frac{1}{\Upsilon_v}\sum_{s=2}^\infty\frac{z^s}{(s-1)!}\int \sum_{m=1}^{s-1} \binom{s-1}{m} \label{eq:105} \\
&\times&  \left [ \prod_{i = 1}^m \prod_{j = m+1}^{s}\psi^v_i\chi^v_j \right ] {\cal B}^{(m,s-m)}_{1...s} d\bm{r}_1...d\bm{r}_{s}. \nonumber
\end{eqnarray}

\subsection{\label{subsec:05c}Correlation Functions of the OSE}

As we shall further witness, the OSE may utilize correlation functions of different types, including those of the GCE-type, but this section will deal with the correlation functions setting the probability of a given configuration of particles in an ensemble, which are analogous to (\ref{eq:013}), and which will be denoted, as it has been previously stated, as $\varrho^{(k)}_{1...k}$. We will also consider here ${\cal F}^{(k)}_{1...k}$ and ${\cal A}^{(m,k)}_{1...m+k}$ developed on their basis.

Let us turn back to Eq. (\ref{eq:086}). As it becomes clear from (\ref{eq:063}), (\ref{eq:090}), the essence of the shift to the OSE is in the passage to the limit (\ref{eq:089}). So, we may set
\begin{equation}
\varrho^{(m)}_{1...m}(z)=\lim_{V\rightarrow \infty}\varrho^{(m)}_{G,1...m}(z,\psi^V),
\label{eq:106}
\end{equation} 
thus, as a counterpart to (\ref{eq:086}) we have
\begin{eqnarray}
 \varrho^{(m)}_{1...m}(z)&=& z^m {\cal B}^{(m,0)}_{1...m} \label{eq:107}  \\
 &+& z^m\sum_{n=1}^\infty \frac{z^n}{n!} \int {\cal B}^{(m,n)}_{1...m+n} d\bm{r}_{m+1}...d\bm{r}_{m+n} \nonumber
\end{eqnarray}
(let us remind the reader that when the window functions are not indicated, integration is performed in the infinite limits). Instead of (\ref{eq:058}) we write
\begin{eqnarray}
{\cal F}^{(k)}_{1...k}&=& z^k{\cal  U}^{(k)}_{1...k}  \label{eq:108}  \\
& +& z^k\sum_{n=1}^\infty \frac{z^n}{n!}\int  {\cal  U}^{(k+n)}_{1...k+n} d\bm{r}_{k+1}...d\bm{r}_{k+n}    \nonumber
\end{eqnarray}
while instead of (\ref{eq:085}) we have
\begin{eqnarray}
&&\mkern -12mu{\cal A}^{(m,k)}_{1...m+k}(z) = {\cal B}^{(m,k)}_{1...m+k} \label{eq:109}  \\
&&\mkern +12mu + \sum_{n=1}^\infty \frac{z^n}{n!} \int {\cal B}^{(m,k+n)}_{1...m+k+n} d\bm{r}_{m+k+1}...d\bm{r}_{m+k+n}. \nonumber
\end{eqnarray}

Obviously, quantities ${\cal F}^{(k)}_{1...k}$ and ${\cal A}^{(m,k)}_{1...m+k}$ comply with relations (\ref{eq:019}) and (\ref{eq:031}) correspondingly, and the passage to the limit can be performed always, because the terms of all series are integrated over the coordinates of localized particles only.

Equations (\ref{eq:107})-(\ref{eq:109}) convincingly show that unlike their GCE counterparts  quantities $\varrho^{(m)}_{1...m}$, ${\cal F}^{(k)}_{1...k}$ and ${\cal A}^{(m,k)}_{1...m+k}$ possess the property of translation invariance and by that reflect the properties of a uniform system.

In particular, the difference between these pairs is significant in the cases where a group of particles $1,...k$ is located at a microscopic distance from the boundaries of the volume set by function $\psi^V$. However, there is another option apart from this case, where free particles are found outside the limits of this volume.

The expressions analogous to (\ref{eq:034}), (\ref{eq:035}), (\ref{eq:037}) shall be written as
\begin{equation}
\int {\cal A}^{(m,k)}_{1...m+k} d\bm{r}_{m+k} =  \frac{\partial{\cal A}^{(m,k-1)}_{1...m+k-1}}{\partial z },
\label{eq:110}
\end{equation}

\begin{equation}
\int{\cal A}^{(m,k)}_{1...m+k} d\bm{r}_{l}...d\bm{r}_{m+k}  
=   \frac{\partial^{m+k-l+1}{\cal A}^{(m,l-m-1)}_{1...l-1}}{\partial z^{m+k-l+1}},
\label{eq:111}
\end{equation}
where $l = m+1, m+2, ..., m+k$, and 
\begin{equation}
\int {\cal A}^{(m,k)}_{1...m+k} d\bm{r}_{m+1}...d\bm{r}_{m+k}  
=  \frac{\partial^{k}}{\partial z^{k}} \left [ \frac{{\varrho}^{(m)}_{1...m}}{z^m} \right ].
\label{eq:112}
\end{equation} 
  
Once again, we remind that the integration should be performed over the coordinates of localized particles only.

Eq. (\ref{eq:110}) is easy to explicitly obtain from (\ref{eq:109}), or from (\ref{eq:034}) by applying the passage to the limit (\ref{eq:089}). Equations (\ref{eq:111}) and (\ref{eq:112}) obviously follow from (\ref{eq:110}).

At $m=1$ (\ref{eq:110}) - (\ref{eq:112}) yield the OSE counterparts of (\ref{eq:025}), (\ref{eq:026}) and (\ref{eq:027}) for ${\cal F}^{(k)}_{1...k}$:
\begin{equation}
\int {\cal F}^{(k)}_{1...k} d\bm{r}_k = z^k  \frac{\partial}{\partial z } \left[ \frac{{\cal F}^{(k-1)}_{1...k-1}}{z^{k-1}} \right],
\label{eq:113}
\end{equation}
\begin{equation}
\int {\cal F}^{(k)}_{1...k} d\bm{r}_l...d\bm{r}_k 
 = z^k  \frac{\partial^{k-l+1}}{\partial z^{k-l+1} } \left[ \frac{{\cal F}^{(l-1)}_{1...l-1}}{z^{l-1}} \right], 
\label{eq:114}
\end{equation}
where $l=2,3...k$. For $l=2$ we shall have
\begin{equation}
\int {\cal F}^{(k)}_{1...k} d\bm{r}_2...d\bm{r}_k  
= z^k  \frac{\partial^{k-1}}{\partial z^{k-1} } \left[ \frac{\varrho}{z} \right].
\label{eq:115}
\end{equation}
 
Again taking (\ref{eq:028}) into account we eventually obtain
\begin{equation}
\int {\cal F}^{(k)}_{1...k} d\bm{r}_2...d\bm{r}_k = z^k \beta \frac{\partial^{k}P}{\partial z^{k} }.
\label{eq:116}
\end{equation}

Eq. (\ref{eq:116}) unlike (\ref{eq:029}) is precise, since from (\ref{eq:107}) it follows that $\varrho$ corresponds to (\ref{eq:068}) and (\ref{eq:028}). Thus, we have obtained the exact analog to the Ornstein-Zernike equation in the form of relations (\ref{eq:115}), (\ref{eq:116}).

Clearly enough, a direct analogue of (\ref{eq:015}) for the OSE is impossible. Indeed, as the environment is limitless, equations cannot contain extensive quantities. The artificial singling out of a certain scope of integration will lead to a forced limitation of the configuration of free particles. Leaping ahead, we shall say that this limitation is eliminated when integration is performed over the coordinates of all particles, like in (\ref{eq:126}).

The adjusted counterpart of (\ref{eq:015}) will be the form (\ref{eq:110}) at $k=1$. Indeed, by developing it in accordance with (\ref{eq:032}),
\begin{equation}
\int [\varrho^{(m+1)}_{1...m+1} -  \varrho^{(m)}_{1...m} \varrho] d\bm{r}_{m+k} = z^{m+1} \frac{\partial }{\partial z }\left[ \frac{\varrho^{(m)}_{1...m}}{z^m} \right],
\label{eq:117}
\end{equation}
we witness that, in the given sense, it corresponds to (\ref{eq:015}). By substituting series (\ref{eq:107}) in (\ref{eq:117}) we can clearly see that this expression yields the recurrence relation (\ref{eq:053}), which, in its turn, will be proved in the following section.

\subsection{\label{subsec:05d}The Use of Generating Functions}

Reconstitute (\ref{eq:090}) to a slightly different form by summing the series over $\chi^v$. We obtain:
\begin{equation}
p^v_m = \frac{1}{m! \Upsilon_v} \int \left [ \prod_{i = 1}^m \psi^v_i \right ] \varrho^{(m)}_{G,1...m}(\chi^v) d\bm{r}_1...d\bm{r}_m, 
\label{eq:118}
\end{equation}
where $\varrho^{(m)}_{G,1...m}(\chi^v)$ is defined by expression (\ref{eq:086}).

Here we face the case where all the coordinates of particles $1,...m$ that are of interest to us lie outside the boundary of the range of assignment of the correlation function set by functions $\chi^v$.

First of all, note the similarity of expression (\ref{eq:118}) with the common term of the distribution of the GCE (\ref{eq:008}). In the following approximation 
\begin{equation}
\varrho^{(m)}_{G,1...m} \approx z^m \exp(-\beta U^{m}_{1...m}) , 
\label{eq:119}
\end{equation}
they correspond to each other  with the accuracy to a normalizing factor, that is, in the limit of low densities when the first term plays the major role in the expansion of a correlation function. (This means, in particular, that for low densities the significance of the surface tension is reduced.)

The condition of normalization yields the expression
\begin{equation}
\Upsilon_v = 1+ \sum_{m=1}^{\infty} \frac{1}{m!} \int \left [ \prod_{i = 1}^m \psi^v_i \right ] \varrho^{(m)}_{G,1...m}(\chi^v) d\bm{r}_1...d\bm{r}_m. 
\label{eq:120}
\end{equation}

By taking its logarithm we obtain
\begin{equation}
\ln{\Upsilon_v} = \sum_{m=1}^{\infty} \frac{1}{m!} \int \left [ \prod_{i = 1}^m \psi^v_i \right ] {\cal F}^{(m)}_{G,1...m}(\chi^v) d\bm{r}_1...d\bm{r}_m. 
\label{eq:121}
\end{equation}

By making use of (\ref{eq:058}) and changing again the order of summing as well as regrouping the products of the window functions we obtain Eq. (\ref{eq:076}). Since we started from (\ref{eq:090}), it means that we proved the recurrence relations (\ref{eq:053}), (\ref{eq:054}) and the class of relations (\ref{eq:056}).

Then, by utilizing (\ref{eq:118}) we average the expression $m(m-1)(m-2)...(m-k+1)$:
\begin{eqnarray}
\left \langle  \frac{m!}{(m-k)!} \right \rangle &=& \frac{1}{\Upsilon_v} \sum_{n=0}^{\infty} \frac{1}{n!} \label{eq:122} \\
&\times& \int \left [ \prod_{i = 1}^{k+n} \psi^v_i \right ] \varrho^{(k+n)}_{G,1...k+n}(\chi^v) d\bm{r}_1...d\bm{r}_{k+n}. 
\nonumber
\end{eqnarray}

Let us divide the series in (\ref{eq:122}) by employing $\Upsilon_v$ as series (\ref{eq:120}). Likewise to (\ref{eq:080}) we have:
\begin{eqnarray}
\left \langle  \frac{m!}{(m-k)!} \right \rangle &=& z^k\sum_{n=0}^{\infty} \frac{z^n}{n!} \label{eq:123} \\
&\times& \int \left [ \prod_{i = 1}^{k+n} \psi^v_i \right ] {\cal A}^{(k,n)}_{G,1...k+n}(\chi^v) d\bm{r}_1...d\bm{r}_{k+n}. 
\nonumber
\end{eqnarray}

By substituting the expression with the window functions $\chi^v$ which is analogous to (\ref{eq:085}) in (\ref{eq:123}) we obtain, via repeating the already habitual procedure of changing the order of summing and regrouping the window functions:
\begin{equation}
\left \langle  \frac{m!}{(m-k)!} \right \rangle = z^k\sum_{t=0}^{\infty} \frac{z^t}{t!} \int \left [ \prod_{i = 1}^k \psi^v_i \right ] {\cal B}^{(k,t)}_{1...k+t} d\bm{r}_1...d\bm{r}_{k+t}. 
\label{eq:124}
\end{equation}

Setting $k=1$ in (\ref{eq:124}) yields the expression for the average particle number $m$:
\begin{equation}
\overline{ m } = z\sum_{t=0}^{\infty} \frac{z^t}{t!} \int  \psi^v_1  {\cal B}^{(1,t)}_{1...t+1} d\bm{r}_1...d\bm{r}_{t+1} , 
\label{eq:125}
\end{equation}
which, evidently, corresponds to (\ref{eq:094}), (\ref{eq:098}). Thus we have found an alternative proof for Eq. (\ref{eq:100}) and, consequently, we proved the class of recurrence relations (\ref{eq:055}).

The summing of series (\ref{eq:124}) in accordance with (\ref{eq:107}), gives:
\begin{eqnarray}
\int \left [ \prod_{i = 1}^k \psi^v_i \right ] \varrho^{(k)}_{1...k} d\bm{r}_1...d\bm{r}_{k} = \left \langle  \frac{m!}{(m-k)!} \right \rangle, 
\label{eq:126}
\end{eqnarray}
which corresponds to (\ref{eq:016}). Thereby, this equation is precise for the correlation functions of the OSE as well as for the GCE. However, as we will further see, the values of average quantities in the OSE are not distorted unlike in the GCE.

\section{\label{sec:06}A System in the Field}

The present article will be dealing with the issue of a system of particles in the field only to the extent, to which it is related to the major topic of discussion that is the formulation of the OSE.

Let us now consider a statistical system located in a field where the scope of the field is far smaller than that of the system and with the field quite far from its boundaries. For such a situation it will sussici to study only the GCE as we will further see. In other words, we will be interested in what happens inside the system and not on its external boundaries.

Furthermore, the foregoing discussion will be based upon the presupposition that function $P(z)$ is analytical and the corresponding series converge. This issue has been raised in a large number of works, e.g. \cite{LebowitzPenrose1968}. So, the present work does not discuss this question and assumes all functions to have analytical character and the corresponding series to be in the state of convergence. 

Apart from that, we shall suggest that the potential of the fields under consideration decays/rises quickly enough for the corresponding integrals to converge inward or outward the body.

\subsection{\label{subsec:06a}Basic Equation}

Consider the configuration integral of a nonuniform system:
\begin{equation}
Z^U_N = \int\limits_{V}^{} \exp(-\beta\sum_{i=1}^N u_i-\beta U^N_{1...N}) d\bm{r}_1...d\bm{r}_N. 
\label{eq:127}
\end{equation}
 
Here $u_i$ is the energy of interaction with the field of the $i$-th particle.   

Let us introduce the functions
\def\bydefn{\stackrel{def}{=}}\def\convf{\hbox{\space \raise-2mm\hbox{$\textstyle      \bigotimes \atop \scriptstyle \omega$} \space}}
\begin{equation}
\varphi_i \bydefn \varphi(\bm{r}_i) = 1 - \exp(-\beta u_i), 
\label{eq:128}
\end{equation}
that depend on the coordinates of  the $i$-th particle, with their advantage being the ability to localize it inside the scope of the field.  The further elimination of $\exp(-\beta u_i)$ from (\ref{eq:127}) yields
\begin{equation}
Z^U_N = \int\limits_{V}^{}(1-\varphi_1)...(1-\varphi_N) \exp(-\beta U^N_{1...N})d\bm{r}_1...d\bm{r}_N. 
\label{eq:129}
\end{equation}

For an analogous system devoid of the field we will obviously obtain:
\begin{equation}
Z_N = \int\limits_{V}^{} \exp(-\beta U^N_{1...N})d\bm{r}_1...d\bm{r}_N. 
\label{eq:130}
\end{equation}

As for the fraction of the partition functions, by integrating over the coordinates of free particles we may write
\begin{equation}
\frac{Z^U_N}{Z_N} = 1 + \sum_{k=1}^N \binom{N}{k} (-1)^k \int\limits_{V}^{} \varphi_1 ...\varphi_k P^{(k)}_{1...k} d\bm{r}_1...d\bm{r}_k.
\label{eq:131}
\end{equation}

Moving on to the distribution functions for an arbitrary set of particles $\varrho^{(k)}_{C,1...k}$ from (\ref{eq:131}) follows:
\begin{equation}
\frac{Z^U_N}{Z_N} = 1 + \sum_{k=1}^N \frac{(-1)^k}{k!}  \int\limits_{V}^{} \varphi_1 ...\varphi_k \varrho^{(k)}_{C,1...k} d\bm{r}_1...d\bm{r}_k.
\label{eq:132}
\end{equation}

Let us turn to the GCE and average Eq. (\ref{eq:132}) over the fluctuation of the particle number for a system without a field. By introducing the value
\begin{equation}
\Xi^U_V = \sum_{N=0}^\infty \frac{z^N Z^U_N}{N!}, 
\label{eq:133}
\end{equation}

which is, evidently, a large partition function for a system in the field, and by changing the order of summing in the right-hand side we shall have 
\begin{equation}
\frac{\Xi^U_V}{\Xi_V} = 1 + \sum_{k=1}^\infty \frac{(-1)^k}{k!}  \int\limits_{V}^{} \varphi_1 ...\varphi_k \varrho^{(k)}_{G,1...k}(\psi^V) d\bm{r}_1 ...d\bm{r}_k.
\label{eq:134}
\end{equation}

Let us take the logarithm of Eq. (\ref{eq:134}).
\begin{eqnarray}
\ln{\Xi^U_V}-\ln{\Xi_V} &=& \sum_{k=1}^\infty \frac{(-1)^k}{k!}\label{eq:135} \\
&\times& \int\limits_{V}^{} \varphi_1 ...\varphi_k {\cal F}^{(k)}_{G,1...k} (\psi^V) d\bm{r}_1...d\bm{r}_k. 
\nonumber
\end{eqnarray}

As it was mentioned before, we are considering the case where the distance from the field scope to the boundaries of the system is large. On these conditions, the comparison of (\ref{eq:058}) with (\ref{eq:108}) makes evident, as a consequence of the local character of Ursell factors, with a very fine accuracy that
\begin{equation}
 {\cal F}^{(k)}_{1...k} =  {\cal F}^{(k)}_{G,1...k}(\psi^V),
\label{eq:136}
\end{equation}
that is why in the frame of these assumptions the following is true:
\begin{equation}
\ln{\Xi^U_V}-\ln{\Xi_V} = \sum_{k=1}^\infty \frac{(-1)^k}{k!}  \int \limits_{V}^{}\varphi_1 ...\varphi_k {\cal F}^{(k)}_{1...k} d\bm{r}_1...d\bm{r}_k.
\label{eq:137}
\end{equation}

This equation is basic for the further consideration of the field related issues. As we can see, the variables here are also separated: $\varphi$ are the functions of the field, whereas Ursell functions belong to a uniform system without a field.

\subsection{\label{subsec:06b}Smooth Field}

Let us set the scale where a field is subject to significant changes as macroscopic. Then, due to the microscopic region of the decay of ${\cal F}^{(k)}_{1...k}$ we may employ (\ref{eq:116}) and write

\begin{equation}
{\cal F}^{(k)}_{1...k}  = z^k \beta \frac{\partial^{k}P}{\partial z^{k} } \prod_{l = 2}^k \delta(\bm{r}_l - \bm{r}_1),
\label{eq:138}
\end{equation}
where $k=2,3...$. (It is necessary to mention again that (\ref{eq:138}) is true for $k$'s that do not reach macroscopic values.)  By substituting (\ref{eq:138}) in (\ref{eq:137}) and taking the integral over the coordinates of all particles but the first one we have:
\begin{equation}
\ln{\Xi^U_V}-\ln{\Xi_V} = \beta \int\limits_{V}^{}\sum_{k=1}^\infty \frac{(-1)^k}{k!} \varphi_1^k z^k  \frac{\partial^{k}P}{\partial z^{k} } d\bm{r}_1,
\label{eq:139}
\end{equation}
where we resorted to Eq. (\ref{eq:028}) to reconstitute the first term of the sum. Summing the Taylor series yields:
\begin{equation}
\ln{\Xi^U_V}-\ln{\Xi_V} = \beta \int\limits_{V}^{}\left [ P(z e^{- \beta u_1}) - P(z) \right ] d\bm{r}_1,
\label{eq:140}
\end{equation}
or, by turning to the $\Omega$-potential
\begin{equation}
\Omega_V-\Omega^U_V =  \int\limits_{V}^{}\left [ P(z e^{- \beta u_1}) - P(z) \right ] d\bm{r}_1
\label{eq:141}
\end{equation}
and, by eliminating the same terms, we write:
\begin{eqnarray}
\Omega^U_V &=&  -\int\limits_{V}^{} P(z e^{- \beta u_1}) d\bm{r}_1 \nonumber \\
 &=& -\int\limits_{V}^{} P[\mu - u(\bm{r})] d\bm{r} = -\int\limits_{V}^{} P(\bm{r}) d\bm{r},
\label{eq:142}
\end{eqnarray}
that is the outcome apprehensible from the point of view of physics, expressing the additivity of thermodynamic quantities of a nonuniform system. Here, we applied the well-know expression for the chemical potential in a field:
\begin{equation}
\mu =  \mu_0 (P) + u(\bm{r}),
\label{eq:143}
\end{equation}
where $\mu_0$ is the chemical potential observed in the absence of a field as a function of pressure.

Let us make use of relation (\ref{eq:142}) to calculate the "`surface tension"' of a boundary of macroscopic thickness. Assume that the potential have a solid core and a smooth boundary. For the sake of simplicity we also presume that it does not turn negative anywhere and that it monotonously falls as it distances away from the field region. We shall further denote the core volume by $V_1$, and that of the transition zone - by $V_t$. Whereas $V_t \ll V_1$, from (\ref{eq:142}) we may write 
\begin{equation}
\Omega = -P(V-V_1-V_t) - A\int\limits_{L_t}^{} P(\bm{r}) dx,
\label{eq:144}
\end{equation}
where $A$ stands for the area limiting the field region, $P$ - is the pressure in the region where the field is absent, $L_t$ - is the length of the transition zone, while $x$ - is the coordinate directed along the gradient of the field. Comparing (\ref{eq:144}) to the standard definition of the $\Omega$-potential for a two-phase system in equilibrium  
\begin{equation}
\Omega = -P(V-V_1-V_t) + \sigma A
\label{eq:145}
\end{equation}
(since the system volume reduced by $V_1+V_t$), we have 
\begin{equation}
\sigma = - \int\limits_{L_t}^{} P(\bm{r}) dx,
\label{eq:146}
\end{equation}
at the same time, the number of surface particles $N_S$ is defined as: 
\begin{eqnarray}
N_S &=& - \frac{\partial \sigma A}{\partial \mu} = A \frac{\partial }{\partial \mu}  \int\limits_{L_t}^{} P(\bm{r}) dx \label{eq:147}\\
 &=& A   \int\limits_{L_t}^{}\frac{\partial }{\partial \mu} P[\mu - u(\bm{r})] dx = A   \int\limits_{L_t}^{} \varrho(\bm{r}) dx. \nonumber
\end{eqnarray}

So, in this case we have $\sigma < 0$, and $N_S > 0$.

However we can include the transition zone to the volume term just as well and, alongside (\ref{eq:144}), come up with an alternative outcome:
\begin{equation}
\Omega = -P(V-V_1) + A\int\limits_{L_t}^{} \left [P - P(\bm{r}) \right ]dx.
\label{eq:148}
\end{equation}

Provided that:
\begin{equation}
\sigma = \int\limits_{L_t}^{} \left [P - P(\bm{r}) \right ]dx
\label{eq:149}
\end{equation}
and
\begin{equation}
N_S = -  A \int\limits_{L_t}^{} [\varrho - \varrho(\bm{r})] dx, 
\label{eq:150}
\end{equation}
where $\varrho$ - is the density in the region where the field is absent. We can see that now it is $\sigma > 0$, and $N_S < 0$.

Evidently, these two points of view have equal rights in this case and are accounted for by two different ways of taking surface effects into consideration: either by adding them as external ones, or subtracting them as internal ones.

We will further see that when we deal with the potential of a hard solid, like in the case of the partition function of the OSE described above, there are no such arbitrary options.

It is interesting that an attempt to set $N_S = 0$ by shifting the boundary between the volume and surface components, like in the case of the phase equilibrium, actually proves to be a bad idea. The point is, relations (\ref{eq:144}) - (\ref{eq:148}) are true for arbitrary $z$ and $T$, since they are not bound by the phase equilibrium curve. The surface obtained on the condition of $N_S = 0$ will, strictly speaking, depend on the value of the chemical potential at a given temperature and thus loose the physical meaning.

It is possible, though, to set to zero the coefficient at the first degree of activity (density). As it can easily be obtained from (\ref{eq:142}) this condition is set by
\begin{equation}
\int\limits_{-\infty}^{l_0} (1 - \varphi_1) dx_1 = \int\limits_{l_0}^{\infty} \varphi_1 dx_1, 
\label{eq:151}
\end{equation}
where the integration in the left-hand side of the equation is performed from the field region to the outside, and in the right-hand one - inward the environment, defining the location of surface $l_0$ that is of interest to us. At that, for the surface tension coefficient and number of surface particles the series expansion in powers of activity starts from the second degree.

So, we have three options of considering of surface effects: one of them starts with a quadratic term of activity, two - with a linear ones. We will discuss them in section \ref{subsec:06d} to provide an example, that is closer to reality.

\subsection{\label{subsec:06c}Sharply Varying Field}

Let us now return to the equation (\ref{eq:137}) and move from the smooth field to the opposite extreme case - the field of a hard solid. The potential energy of particles may be written 
\begin{equation}
u(\bm{r}_i) = u_i = 
	 \left\{ 
			\begin{array}{ll} 
         +\infty & (\bm{r}_i \in v)\\   
         0 & (\bm{r}_i \notin v).
     	\end{array}  
		\right.
		\label{eq:152}
\end{equation}

At the same time, functions $\varphi_i$ (\ref{eq:128}) take the form of window functions (\ref{eq:001}):
\begin{equation}
\varphi_i \rightarrow \psi^v_i.
\label{eq:153}
\end{equation}

Let us assume that this body occupies the macroscopic volume $v$ inside $V$. As it has already been stated, we presuppose that $v \ll V$ and the body is far enough from the boundaries of volume $V$.  Having set so, we can eliminate the sign $V$ in the integrals over the coordinates of particles from (\ref{eq:137}) and consider the integration to be performed over all space: 
\begin{equation}
\ln{\Xi^U_V}-\ln{\Xi_V} = \sum_{k=1}^\infty \frac{(-1)^k}{k!}  \int \left [ \prod_{l = 1}^k \psi^v_l \right ] {\cal F}^{(k)}_{1...k} d\bm{r}_1...d\bm{r}_k.
\label{eq:154}
\end{equation}

So far, this equation does not allow us to perform integration according to (\ref{eq:116}) due to a distortion occurring at the boundary of volume $v$.
For example, when the first particle is at the distance of a "`correlation length"' from the boundary then, in order for (\ref{eq:116}) to fulfill, the integration over the coordinates of remaining particles shall be carried out over all space, and shall not be limited, like in this case, by functions $\psi^v_l$. In order to employ formula (\ref{eq:116}) let us reconstitute the integrals in (\ref{eq:154}) in the way similar to (\ref{eq:066}):
\begin{eqnarray}
\ln{\Xi^U_V}&-&\ln{\Xi_V} = \sum_{k=1}^\infty \frac{(-1)^k}{k!} \int \psi_1^v {\cal F}^{(k)}_{1...k} d\bm{r}_1...d\bm{r}_k \label{eq:155}  \\
&+& \sum_{k=2}^\infty \frac{(-1)^k}{k!} \int \psi_1^v \left (\left [ \prod_{l = 2}^k \psi^v_l \right ]
 - 1  \right ){\cal F}^{(k)}_{1...k} d\bm{r}_1...d\bm{r}_k. \nonumber
\end{eqnarray}

This treatment will ultimately be analogous to the one given above for the OSE, however, we will carry it out at the level of correlation functions instead of that of factors.

By taking into account (\ref{eq:116}), the first sum yields $\beta v[P(0) - P(z)]$,  and in the second one we may express $\psi_i^v$ for $i$=2,3... through $\chi_i^v$ (\ref{eq:069}).

We will thus obtain:
\begin{eqnarray}
&\ln{\Xi^U_V}&-\ln{\Xi_V}=\beta v[P(0) - P(z)]  \label{eq:156} \\ 
&+& \sum_{k=2}^\infty \frac{(-1)^k}{k!}\int \psi_1^v \left [\prod_{l = 2}^k(1-\chi_l^v) - 1  \right ]{\cal F}^{(k)}_{1...k} d\bm{r}_1...d\bm{r}_k.\nonumber  
\end{eqnarray}

In accordance with our supposition of the analyticity of $P(z)$, we assume that $P(0) = 0$. Once again, as it can be easily seen, the structure of integrands in the right-hand side of (\ref{eq:156}) is such that at least one pair of particles always lies at the different sides of the field boundary. At the same time, due to the local properties of ${\cal F}^{(k)}_{1...k}$ the integrals are defined, like in (\ref{eq:070}), by the region close to the surface. We may once again take the integral along the boundary surface over the corresponding coordinates of the first particle, the integrands being independent from them in this case. Upon making these transformations and eliminating the same terms, we write:
\begin{equation}
\Omega^U_V = -(V-v)P(z,T)  +a \sigma (z,T),
\label{eq:157}  
\end{equation}
where $a$ - is the area of the surface of a hard solid submerged in a fluid,
\begin{equation}
P(z,T) =\varrho k_BT -k_BT \sum_{k=2}^\infty \frac{(-1)^k}{k!} \int {\cal F}^{(k)}_{1...k} d\bm{r}_2...d\bm{r}_k, \label{eq:158}  
\end{equation} 
and 
\begin{eqnarray}
\sigma (z,T) &=&  k_BT \sum_{k=2}^\infty \frac{(-1)^k}{k!}
\label{eq:159} \\ 
&\times&  \int \psi_1 \left [1 - \prod_{l = 2}^k(1-\chi_l) \right ]{\cal F}^{(k)}_{1...k} dx_1 d\bm{r}_2...d\bm{r}_k, \nonumber
\end{eqnarray}
where $x_1$ - is the coordinate of the first particle perpendicular to the boundary surface. 

Two equations (\ref{eq:158}), (\ref{eq:159}) are equivalent to the pair (\ref{eq:068}), (\ref{eq:074}) in the cases when the latter ones are true. This can be easily proved by substituting expansion (\ref{eq:108}) to (\ref{eq:158}), (\ref{eq:159}) and changing the order of summing. 

The expressions analogous to (\ref{eq:158}), (\ref{eq:159}) for a multicomponent environment may prove effective in the case of employing the Coulomb interaction of particles.

Relation (\ref{eq:157}) has the same form as it does for a two-phase system in equilibrium, with the exception of the fact that $\sigma$ - the surface tension coefficient - started depending on two variables. Indeed, upon introducing the field of an infinitely positive potential, the system volume became equal to $V-v$ and we obtain (\ref{eq:157}).

Note that, as it has been previously said, we cannot move the volume boundary in the case of the field of a hard solid, unlike that of a "`thick"' surface. The boundary is rigidly set, whereas the volume and surface terms have different functional relationships with the activity: the linear relation is opposed to the square one at low densities. Thus, in this case, there is no arbitrariness in the sign and value of $\sigma$ and $N_S$.

So, the expression for $\sigma(z,T)$, occurring in the expression for the partition function of the OSE corresponds ideally to the surface tension observed at the boundary between a hard solid body and a fluid. This correspondence is by no means coincident and will be later discussed.

\subsection{\label{subsec:06d}Actual Solid}

Let us once more go back to Eq. (\ref{eq:137}) and examine it for the case of the potential of a real solid body. Obviously, we will consider it constant, so that the situation corresponds to the metastable state, once again allowing us to define the surface tension as a function of two variables.

We will first discuss the model case of the field independent from displacements along the surface of the body and then we will generalize it.

Let us separate (\ref{eq:137}) into the volume and surface terms like in (\ref{eq:155}). We shall obtain:
\begin{equation}
\Omega = -PV +P\int\limits_{V}^{} \varphi_1 d\bm{r}_1 + \sigma_u A,
\label{eq:160}
\end{equation}
where we introduced the notation
\begin{eqnarray}
\sigma_u (z,T) &=&  k_BT \sum_{k=2}^\infty \frac{(-1)^k}{k!}
\label{eq:161} \\ 
&\times&  \int \varphi_1 \left [1 - \prod_{l = 2}^k(1-\theta_l) \right ]{\cal F}^{(k)}_{1...k} dx_1 d\bm{r}_2...d\bm{r}_k, \nonumber
\end{eqnarray}
where $x_1$ - is again the coordinate directed along the gradient of the field  and $\theta$ is defined as
\begin{equation}
\theta_l = 1 - \varphi_l = \exp(-\beta u_l).
\label{eq:162}  
\end{equation}

Eq. (\ref{eq:161}), in the same way to the cases investigated above, may be reconstituted by employing (\ref{eq:108}). Then we have:
\begin{eqnarray}
\sigma_u (z,T) &=&  k_BT \sum_{k=2}^\infty \frac{z^k}{k!}
\label{eq:163} \\ 
&\times& \int   \theta_1  \left [1 - \prod_{l = 2}^k(1-\varphi_l) \right ]{\cal U}^{(k)}_{1...k} dx_1 d\bm{r}_2...d\bm{r}_k. \nonumber
\end{eqnarray}

The first terms of (\ref{eq:163}) were calculated in a series of works by the authors of \cite{SokolowskiStecki1981} on the basis of the topological approach from \cite{Bellemans1962}, though extending it beyond the limits of the potential of a hard solid body.

Let us reconstitute (\ref{eq:160}) to
\begin{eqnarray}
\Omega = -P(V-V') &-& PA\int\limits_{-\infty}^{x'}(1 - \varphi_1) dx_1 \label{eq:164} \\
 &&+ PA\int\limits_{x'}^{\infty}\varphi_1 dx_1 + \sigma_u A,\nonumber
\end{eqnarray} 
where $x'$ is some arbitrary point located inside the transition layer or in its vicinity, that defines volume $V'$. 

We can observe the existence of a certain degree of arbitrariness in distinguishing between volume and surface terms. As a matter of fact, this arbitrariness can only exist within the range of the transition layer, i.e. $x'$ is situated inside it. It is easy to see, indeed, that in the opposite case the issue is but a banal compensation of two identical terms of different signs. Thus, like in the case with a "`thick"' surface, we obtain at least three options for taking surface effects into account:
\begin{equation}
\Omega = -P(V-V_1-V_t) - PA\int\limits_{L_t}^{} (1- \varphi_1) dx_1 + \sigma_u A,
\label{eq:165}
\end{equation}
\begin{equation}
\Omega = -P(V-V_1) + PA\int\limits_{L_t}^{} \varphi_1 dx_1 + \sigma_u A,
\label{eq:166}
\end{equation}
\begin{equation}
\Omega = -P(V-V_0) + \sigma_u A,
\label{eq:167}
\end{equation}
where the denotations are the same as in (\ref{eq:144}), and $V_0$ - is the body volume delimited by the surface set by condition (\ref{eq:151}). 

Option (\ref{eq:165}) attributes surface effects to a solid body, whereas (\ref{eq:166}) - ascribes them to a fluid. Option (\ref{eq:167}) sets linear surface terms to zero by making use of property (\ref{eq:151}). 

Generally speaking, option (\ref{eq:165}) seems to be preferable when real adsorption is observed, since it is in agreement with the ordinary notion of particles adsorbed on the surface of a solid body. As we can see, in the first order term of $z$ it always shows a positive value of the surface particles number.

Option (\ref{eq:167}) is not suited for the case of a real adsorption, because it attributes the molecules adsorbed on the surface to the volume of a fluid, being in this case an apparently artificial approach. Indeed, the presence of a strongly expressed negative potential in the vicinity of the surface, as can be easily seen from (\ref{eq:151}), shifts the position of surface $l_0$ inward the body. 

This option is helpful when the surface potential is close to the hard solid one, or, in speaking in general terms, when adsorption is absent.

Option (\ref{eq:166}) is the most exotic one; however, it can possibly also be efficient in the case of potentials of unusual form. 

Like in the case of a "thick" surface, the option with the condition $N_s = 0$ is inappropriate - for the same reasons.

In the case, when we are able to employ (\ref{eq:138}) from (\ref{eq:161}) we obtain
\begin{equation}
\sigma_u (z,T) =  \int\limits_{-\infty}^{\infty}\left [ \theta_1 P(z)- P(\theta_1 z)  \right ] dx_1,
\label{eq:168}
\end{equation}
which naturally leads, when substituting it in (\ref{eq:165}), (\ref{eq:166}), to (\ref{eq:144}), (\ref{eq:148}). Obviously, the limits of integration in (\ref{eq:168}) can be restricted to the transition zone.

Now let us consider the situation where the field depends on displacement along the surface. We shall assume that the body is a crystalline solid and thus have the surface structure recurring at some periodical intervals. 

Examining the integrals over the coordinates of the first particle that are parallel to the surface obviously yields their averaging over such a surface unit cell. Its linear dimensions can significantly supersede those of a volume cell at certain structural orientations of the crystal in relation to the surface.

So, (\ref{eq:164}) takes the form
\begin{eqnarray}
\Omega = -P(V-V') &-& PA\int\limits_{-\infty}^{x'}(1 - \bar{\varphi_1}) dx_1 \label{eq:169} \\
 &&~~~+ PA\int\limits_{x'}^{\infty}\bar{\varphi_1} dx_1 + \sigma_{\bar{u}} A,\nonumber
\end{eqnarray}
where $\bar{\varphi_1}$ - is the value averaged over the surface cell at a set depth of a transition zone, and for $\sigma_{\bar{u}}$ we have:

\begin{eqnarray}
\sigma_{\bar{u}} (z,T) &=&  k_BT \sum_{k=2}^\infty \frac{(-1)^k}{k!}
\label{eq:170} \\ 
\times \int &dx_1& \left\langle\varphi_1 \int  \left [1 - \prod_{l = 2}^k(1-\theta_l) \right ] {\cal F}^{(k)}_{1...k}  d\bm{r}_2...d\bm{r}_k\mkern -8mu\right\rangle_{cell}, \nonumber
\end{eqnarray}
where the triangular brackets also stand for the averaging over the surface cell at a set depth of a transition zone. We should remind our reader that ${\cal F}^{(k)}_{1...k}$ is not directly subject to averaging, as they belong to the unperturbed environment.

The counterpart for (\ref{eq:163}) will be
\begin{eqnarray}
\sigma_{\bar{u}} (z,T) &=&  k_BT \sum_{k=2}^\infty \frac{z^k}{k!}
\label{eq:171} \\ 
\times \int& dx_1& \left\langle \theta_1 \int  \left [1 - \prod_{l = 2}^k(1-\varphi_l) \right ]{\cal U}^{(k)}_{1...k}  d\bm{r}_2...d\bm{r}_k\mkern -8mu\right\rangle_{cell}, \nonumber
\end{eqnarray}
where, once again, averaging is performed over the surface cell at a set depth.

Equations (\ref{eq:151}),(\ref{eq:165}) - (\ref{eq:167}) maintain their forms, with $\varphi_1 \rightarrow \bar{\varphi_1}$ and $\sigma_u \rightarrow \sigma_{\bar{u}}$ being substituted.  

In the case, where the contact surface contains crystal faces differently oriented in relation to the crystal axes, the surface terms of expression (\ref{eq:169}) should be summed over these faces.

\section{\label{sec:07}Discussion}

\subsection{\label{subsec:07a}The Implication of the Surface Term for the OSE}

Eq. (\ref{eq:075}) for the probability of cavity formation, which defines the partition function of the OSE, is in agreement with the general expression for the probability of a fluctuation:
\begin{equation}
p_0 \propto \exp{(-\beta R_{min})},
\label{eq:172}  
\end{equation}
\cite[p.339]{landaulifshitz1985}, where $R_{min}$ - is the minimal work done on eliminating the fluctuation. 

Indeed, as the chemical potential for the particles located outside of the field does not change upon a local introduction of a field according to (\ref{eq:143}), then we may assume that, in this case, the process of creation/elimination of a fluctuation is carried out at a constant chemical potential and temperature, i.e. the natural variables of the $\Omega$-potential. Thus, the minimal work on forming a cavity is equal to the alteration of the $\Omega$-potential:
\begin{eqnarray}
R_{min} = \Omega^U - \Omega &=& [-P(V-v) + \sigma a] - [-PV] \nonumber \\
 &=& vP(z,T)  +a \sigma (z,T). \label{eq:173}  
\end{eqnarray}

As we are interested in the fluctuation that causes density to change in a stepwise way, this process of creating the fluctuation must be carried out by a hard solid. It is here where we find the meaning of the correspondence existing between the surface tension coefficient in the OSE and that for the boundary surface of a hard solid and a fluid.

\subsection{\label{subsec:07b}The Difference between the OSE and the GCE  }

The core difference between the two ensembles is in surface effects, that can be observed in (\ref{eq:063}). In the case where the system does not interact with the environment, $\Xi_V = \Xi_v \Xi_{V-v}$ and this equation is deduced to a banal one 
\begin{equation}
\ln{\Upsilon_v} =  \lim_{V\rightarrow \infty} \left [ \ln{(\Xi_v \Xi_{V-v})} - \ln{\Xi_{V-v}} \right ] = \ln{\Xi_v},
\label{eq:174}  
\end{equation}
demonstrating the identity of the ensembles.  

At $v=V$ from (\ref{eq:061}) we obtain
\begin{equation}
\ln{P^V_0} = -\ln{\Xi_V},
\label{eq:175}  
\end{equation}
just like it should be from the point of view of an ordinary treatment of the GCE.

However, the differences are not connected to the presumed lack of surface terms in the GCE; they are due to fact that such terms have another form. Indeed, by applying the operation of singling out the surface term to (\ref{eq:062}) we may write:
\begin{eqnarray}
\ln{\Xi_V} &=& \sum_{t=1}^\infty \frac{z^t}{t!} \int \psi^V_1 {\cal U}^{(t)}_{1...t} d\bm{r}_1...d\bm{r}_t \label{eq:176} \\
&-& \sum_{t=2}^\infty \frac{z^t}{t!} \int \psi^V_1 \left [1 - \prod_{i = 2}^{t} (1 - \chi^V_i) \right ] {\cal U}^{(t)}_{1...t} d\bm{r}_1...d\bm{r}_t.\nonumber
\end{eqnarray}

By comparing (\ref{eq:176}) with (\ref{eq:067}), (\ref{eq:073}) we witness that
\begin{equation}
\Xi_V = \exp{\beta [ VP(z,T)  - A\sigma (z,T) ]},
\label{eq:177}
\end{equation}
where we resorted to the substitution $\psi \leftrightarrow \chi$ in the last term of (\ref{eq:176}), which always remains possible due to the spatial symmetry of the task.

Let us write the formulas (\ref{eq:072}), (\ref{eq:097}), (\ref{eq:100}) of the OSE once again for the sake of comparison:
\begin{equation}
\Upsilon_V = \exp{\beta [ VP(z,T)  + A\sigma (z,T) ]},
\label{eq:178}
\end{equation}
\begin{equation}
\frac{\partial\ln{\Upsilon_V}}{\partial \ln{z}} = N_b - N_s = \overline{N} - N_s,
\label{eq:179}
\end{equation} 
and, by differentiating (\ref{eq:177}), those of the GCE:
\begin{equation}
\frac{\partial\ln{\Xi_V}}{\partial \ln{z}} = N_b + N_s = \overline{N}.
\label{eq:180}
\end{equation} 

(The last equality in (\ref{eq:180}), as is known, can be easily obtained by differentiating (\ref{eq:010}) over the chemical potential.)

Thus, the distributions of the OSE and GCE contain the surface terms of different signs. We will not be examining this issue profoundly right now. We would like to briefly note, though, that the reverse sign of the surface term in (\ref{eq:177}) is no coincidence, and that the GCE, being a set of closed systems, cannot reproduce the open system in all its completeness.

Now we can turn our attention back to (\ref{eq:029}). From (\ref{eq:016}) and (\ref{eq:180}) follows that
\begin{equation}
\int \psi^V_1 \varrho^{(1)}_{G,1} (\psi^V) d\bm{r}_1 = \overline{N} = N_b + N_s,
\label{eq:181}
\end{equation}
therefore, we see that $\varrho^{(k)}_{G,1...k}$ contains surface terms and that is the exact reason for (\ref{eq:029}) to not be fulfilled when the first particle is located close to the boundary of volume $V$.

As a comparison, from (\ref{eq:126}) and (\ref{eq:179}) we have for the OSE:
\begin{eqnarray}
\int \psi^V_1 \varrho^{(1)}_{1} d\bm{r}_1 = \overline{N} = N_b, 
\label{eq:182}
\end{eqnarray}
which is due to the exact satisfaction of (\ref{eq:116}) independently of the position of the first particle.

Eq. (\ref{eq:175}) demonstrates the lack of scalability of the GCE or, in other words, it shows that this distribution does not match the conformity criterion.

The GCE does not scale because in the case of $v=V$ it yields the expression different from the case where $v<V$, which is evident from the comparison of (\ref{eq:061}) and (\ref{eq:175}). We shall further note that the volume term, obviously, maintains its form and, thus, this part of the GCE is scalable.

Indeed, considering expression (\ref{eq:058}), leads to the understanding that the presence of the window functions $\psi^V_l$ distorts the correlation functions for those configurations when a group of particles from the 1st to $k$th is located in the vicinity of the boundary of volume $V$. It happens as a consequence of the integrals of the Ursell factors do not reach in this case their extreme values as in integrating over the whole space or the whole area where they are different from zero.

Clearly, the correlation functions in the uniform environment should possess the property of translation invariance, which is not fulfilled for GCE in the vicinity of the boundaries of volume $V$. In order to eliminate this distortion we should pass to limit (\ref{eq:089}), and that is exactly what happens particularly in equations (\ref{eq:063}), (\ref{eq:090}). It is by liquidating this distortion that we can obtain the right expressions  for the surface tension in the OSE.

Let us also note, that such a transition is impossible if we make use of relations like (\ref{eq:015}), incorporating extensive values. However, they can always be deduced to form (\ref{eq:024}), by, for example, employing (\ref{eq:020}), and after that we turn to limit (\ref{eq:089}).

\subsection{\label{subsec:07c}Small Fluctuations}

Logical reasoning tells us that the value of surface terms should decrease for the terms of the distribution of the OSE approaching the average quantities. It would be strange indeed if the probability to find an average number of particles in a certain volume contained the surface. As it has already been said such compensation does take place.

First of all, the very equation (\ref{eq:100}) indicates that. Since, at small values of the particle number in a given volume, the probability depends on the size of the limiting surface, which is clear from (\ref{eq:075}), and since the average value does not depend on it, the only option would be compensating the surface values for the terms of the distributions with larger $m$'s.

Second, according to the calculations, we observe the strict compensation of the surface terms in the vicinity of average values for the first terms of the expansion over activity.

\section{\label{sec:08}Summary}

The article consists of two parts: the description a new statistical ensemble and the consideration of surface effects for a system in the field that corresponds to a solid body. Although, the latter part is auxiliary it contains some results that are of their own individual value. 

\vspace{20 pt} 

New Ensemble
\begin{enumerate}
	\item We introduce a new statistical ensemble named the OSE (open statistical ensemble), whose major feature is in correctly taking into account the surface terms for an open system, which leads to substituting the Boltzmann factor in the configuration integrals of the grand canonical ensemble (GCE) with  the correlation function of a specific form (\ref{eq:118}).
	\item The partition function of the OSE in its expanded form is defined by expression (\ref{eq:064}), which includes volume and surface terms (\ref{eq:072}). The analogous equation for the distribution of the OSE is given by the sum (\ref{eq:090}), whose first term represents GCE.  
	\item Unlike in the case of a two-phase system in equilibrium, the surface tension coefficient for the OSE depends on two variables: pressure as well as temperature (\ref{eq:074}).
	\item The expression for the partition function of the OSE conforms with the thermodynamic approach (\ref{eq:173}).
	\item Thermodynamic and statistical relations of the OSE (section \ref{sec:05}) are ensured by the recurrence relations (\ref{eq:051}) - (\ref{eq:056}) of a new class of correlation functions ${\cal B}^{(m,k)}_{1...m+k}$ and by the use of generating functions (subsection \ref{subsec:05d}).
	\item Functions ${\cal B}^{(m,k)}_{1...m+k}$ (\ref{eq:045}) generalize the notions of the Boltzmann  (\ref{eq:041}) and Ursell (\ref{eq:042}) factors and contain them as extreme cases (\ref{eq:048}), (\ref{eq:049}).
	\item Functions ${\cal B}^{(m,k)}_{1...m+k}$ and ${\cal A}^{(m,k)}_{1...m+k}$ contiguous to them are yielded by fractional generating functions (\ref{eq:046}), (\ref{eq:038}).	
	\item The correct consideration of the integrals of Ursell functions for the OSE results in obtaining the exact analogs for the Ornstein-Zernike equation of higher orders (\ref{eq:116}). (\ref{eq:112}) is a further generalization of this relation.	
	\item The basic equations of the OSE can be obtained on the level of factors as well as on the level of correlation functions thus alleviating the problems related to the convergence of series in density. For instance, we can thus yield the expressions for the surface terms (\ref{eq:161}) and (\ref{eq:163}).
	\item The mean particle number for the OSE does not contain surface terms (\ref{eq:100}), which indicates them being compensated by the high-order terms of distribution.
	\item The GCE contains the surface terms, whose sign is contrary to that of the OSE (\ref{eq:177}), (\ref{eq:178}).
	\item As a consequence, many expressions of the GCE feature inaccuracies related to the surface effects (\ref{eq:029}), (\ref{eq:181}).

\end{enumerate}	

\vspace{10 pt} 

A System in the Field
\begin{enumerate}
	\item The investigation of the system of particles in a field of force of general type proves that the surface terms of the OSE (\ref{eq:159}) really correspond to the surface tension observed at the boundary separating a hard solid from a fluid. This rigid character of the solid is explained by the fact that from the viewpoint of probability distribution we are interested in the configurations abruptly breaking off in density.
	\item The consideration of a solid body as an external field is a productive approach, serving as an alternative to the equilibrium one (a nonvaporizing or insoluble body).
	\item The statistical approach gives a closed expression for the $\Omega$ - potential of a nonuniform system (\ref{eq:142}) in the case of the macroinhomogeneity, which has a clear physical meaning. 
	\item In general, the volume term, adsorption and surface tension cannot be separated with outmost precision (\ref{eq:160}), and it is expedient to interpret them on the basis of the form of the potential of the body and fluid particles interaction (\ref{eq:165}) - (\ref{eq:167}). 
	\item In particular, depending on the situation, surface terms can start with both a linear and quadratic term of activity (\ref{eq:165}), (\ref{eq:167}).
	\item The condition $N_s = 0$ is, by all appearances, always adverse for the case being considered, since the position of the separating surface (\ref{eq:151}) starts depending on pressure.
	\item Surface terms, like in the case of a sharply varying potential, can be calculated in its general form for a smooth field (\ref{eq:168}) as well as for a real one (\ref{eq:163}), including the case of the surface inhomogeneity (\ref{eq:171}).
	\item In the case of a smooth field the series for the surface term converges to a compact form (\ref{eq:168}).

\end{enumerate}

\begin{acknowledgments}
We would like to thank Prof. V.G. Zemlyanukhin for numerous precious discussions and Prof. E.V. Garcia Melijov for help with the translation. 
\end{acknowledgments}

\end{document}